\documentstyle[12pt,amsfonts,epsfig]{article}

\def\Tr{{\rm Tr}}
\def\k{{k_F}}
\def\bfm#1{{\mbox{\boldmath $#1$}}}

\begin{document}

\title{The longitudinal and transverse responses \\
in the inclusive electron scattering: \\
a functional approach}
\author{R.Cenni, F. Conte and P. Saracco\\
Dipartimento di Fisica dell'Universit\`a di Genova \\
 Istituto Nazionale di Fisica Nucleare, Sezione di Genova, \\
 via Dodecaneso, 33 - 16146 Genova (Italy)}
\maketitle
\begin{abstract}
The splitting between the charge-longitudinal and spin-transverse 
responses is explained in a model whose inputs are the effective 
interactions in the particle-hole channels in the frame of the first 
order boson loop expansion. 
It is shown that the interplay between $\omega$-meson exchange and box 
diagrams (two-meson exchange with simultaneous excitation of one or two 
nucleons to $\Delta$'s) mainly rules the longitudinal response, while
in the transverse one the direct $\Delta$ excitations almost cancel the 
one-loop correction and the response is mainly governed by the 
$\rho$-meson rescattering inside the nucleus. It is also shown that a small 
variation in the nuclear densities may explain the observed 
discrepancies between different nuclei.
\end{abstract}

\newline
Preprint GEF-TH-2/97

\newpage

\section{Short introduction\label{sect1}}

We will be concerned in this paper with the separation between longitudinal 
and transverse response in the inclusive electron scattering off nuclei.
We shall present here a microscopic approach which naturally displays 
a depletion of the longitudinal and an enhancement of the transverse 
response.

Before entering the details of the model, some introductory words are 
required both concerning the experimental situation and the present 
status of the theoretical approaches -- and we shall do that in sect. 
\ref{sect2} -- and the philosophy that provided us the guideline in 
constructing our theoretical frame. The latter topic, together with the 
modifications the phenomenology suggested in our line of thought, is  resumed 
in sect.~\ref{sect3}.

Next the bosonic loop expansion formalism, although already described in 
\cite{CeSa-94}, will be resumed, for ease of the reader, in sect.~\ref{sect4}.
There particular emphasis is put on the presence of $\Delta$'s inside the 
diagrams.

Section \ref{sect5} describes the mean field approach. While in the 
charge longitudinal channel this approximation level is too poor both for 
its disagreement with the data and for its poor dynamical content, 
in the transverse one it is able to provide us information about the effective 
interaction in the isovector spin-longitudinal effective interaction, that is 
one of the phenomenological inputs of our model, and this because of
an accidental almost complete cancellation of the 1-loop corrections
in this channel.

In sect. \ref{sect6} the one-loop corrections are described and it is 
shown how the presence of the $\Delta$'s can distinguish the response in the 
two channels.

Finally in sect. \ref{sect7} our conclusions are drawn.

\section{The experimental and theoretical situations\label{sect2}}

The experimental outcomes in the quasi-elastic peak (QEP) region are at 
present still controversial both on the experimental and theoretical point 
of view. Starting from the usual relation
\begin{equation}
\frac{d^2\sigma}{d\Omega d\epsilon}=\sigma_M\left\{v_LR_L(q,\omega)+v_T
R_T(q,\omega)\right\}
\label{lt1}
\end{equation}
the Saclay experimentalists \cite{Me-al-84,Me-al-85} where   
first able to perform the 
Rosenbluth separation, thus getting both $R_L$ and $R_T$. The 
outcomes are known: the longitudinal response, to whom a great amount of 
work has been devoted in these last years (see for instance ref. 
\cite{BoGiPa-93} for a wide review of the current literature), came out to be 
drastically quenched with respect to the Free Fermi Gas (FFG) model, 
while the transverse one was remarkably increased,
a fact, this last, which instead has been less emphasized.

The first difficulties with this Rosenbluth separation came from the 
non-fulfillment of the Coulomb sum rule and the connected problem of the 
so-called missing strength. In fact, the integrated 
longitudinal response, expected to provide the nuclear charge, was 
quenched by  a, say, 10\% in case of Carbon but by a 40\% in the case of 
Calcium. More recent data from Saclay, taken on Lead, showed and even 
enforced the same trend, the sum rule being reduced to about a 40\% of the 
total charge~\cite{Zg-al-94}.

Few years ago the Rosenbluth separation has also been performed on the 
data taken in two different experiments at Bates and the longitudinal 
responses have been published \cite{Ya-al-93}. The quenching of the sum rule 
for the $^{40}Ca$ turns out to be, in this case, of about a 10\%, in sharp 
contrast with the Saclay data. 

Very recently, by reviewing the world data of inclusive electron scattering,
Jourdan~\cite{Jo-95} outlined that the 
Rosenbluth separation is not  free from theoretical ambiguities, and 
other are introduced in deriving the sum rule:
just to exemplify, the one-photon approximation is questionable 
for heavy nuclei and the distortion of the outgoing electron must be 
correctly accounted for, before separating the longitudinal and transverse 
channels; furthermore, relativity prevents us to define the Coulomb sum 
rule in a natural way \cite{Ba-al-94} and some ad hoc relativistic corrections  
are required; moreover the contributions from experimentally
unexplored or physically unaccessible regions in the high energy tail of the 
response must be correctly accounted for.
By considering all these topics (and others neglected here for sake of 
brevity) Jourdan showed that the corrected sum rule derived from world set of 
data is compatible with $Z$ within a 1\% incertitude.

This outcome is, in principle, not strongly contradictory with the Saclay 
results: the nonrelativistic Coulomb sum rule in fact properly reads
\begin{equation}
\int\limits_0^\infty\frac{RE_L(q,\omega)}{G_E^2(q^2)}\simeq 
S_L(q)=Z+\frac{2Z}{\rho}g(q)\;,
\label{lt2}
\end{equation}
$g(q)$ being the pair correlation function, and the outcome of \cite{Jo-95} 
only states that $g(q)$ is compatible with 0 at, say, $q=$570 MeV/c
(the largest transferred momentum examined there). 
In ref. \cite{Jo-96} it is shown indeed that at lower momenta still a 
sizeable quenching (even if less pronounced than in 
\cite{Me-al-84,Me-al-85}) survives.

Waiting for new experimental data that would (hopefully) solve the 
present contradiction, we followed, as a guideline, the Saclay data, 
plotting however, if available, also the data from Bates and those 
derived from the world data set according to \cite{Jo-95,Jo-96}.

To better explain this choice, let us anticipate a little the topics of 
sects. \ref{sect3} and \ref{sect4}: in order to determine a form for the 
effective interaction and then to perform numerical calculations we need 
to refer to some phenomenological inputs. In this respect Saclay data 
display a wide range of transferred momenta both in the longitudinal and 
transverse channel, while those from Bates are limited, up to our 
knowledge, to the longitudinal response only and cannot be used to set 
our inputs. We will see, instead, that the data of refs. 
\cite{Jo-95,Jo-96} can also be obtained by slightly alter our 
parametrization, but without destroying
the physical picture 
we are going to derive in the following.
The physical interpretation however will change significantly.
We defer to the conclusion a more complete discussion about this point.

Coming now to the theoretical side of the problem, to our knowledge the 
longitudinal response has been extensively investigated,  while the 
transverse still requires much attention.

Concerning the former, all theoretical approaches, based on a variety of 
dynamical models, agree in 
providing a more or less pronounced depletion of the QEP in the 
longitudinal channel \cite{FaPa-87,CoQuSmWa-88,Ba-al-95,VaRyWa-95}. The 
different physical inputs and even the different languages present in 
the quoted (and many other) papers seem however to prevent a clear 
comparison between them.

The transverse response has been instead less 
investigated. The common trend seems to be the need of including at least
two particles-two holes excitations to describe the QEP and dip region together
\cite{VaRyWa-95,AlErMo-84,AmCoLa-94}. In particular ref. \cite{VaRyWa-95} 
seems to be (up to our knowledge) the only attempt to explain 
simultaneously the two responses within the same frame, namely that of 
continuum RPA plus two-body currents. Still, both responses 
seem to be slightly overestimated. Further, as in finite nucleus 
calculations, the use of the phenomenological Skyrme interaction obscures the 
link with the underlying dynamics. On the other hand the dynamically richest 
approach is provided in our opinion by the FHNC calculations \cite{FaPa-87} but 
in this case the possibility of exploiting the transverse response is 
precluded by the difficulty (only for the moment, we hope 
\cite{CeFa-93,WaCeFaFa-96,Fa-96-t}) 
of embedding pions and $\Delta$'s in a variational 
approach, which in turns forbids the inclusion in the calculation, at 
least in a natural way,  of the 
Meson Exchange Currents (MEC).

Furthermore, the exact preservation of the sum rules, gauge invariance and 
other general theorems is not in general ensured by the usual many-body 
techniques. Thus some more efforts in explaining the two responses are
still required.

\section{The theoretical frame \label{sect3}}

Having shortly revisited the experimental and theoretical situation, let us
spend a few words about the philosophy underlying our approach 
\cite{CeSa-94,CeCoSa-96}.

Our starting point in \cite{AlCeMoSa-87} was to build up a 
well-behaved approximation scheme for a system of nucleons and pions.

By one side "well-behaved scheme" means an expansion of the physical 
quantities
we are looking for, whatever they are, in powers of a given parameter. This 
because the unicity of the Taylor expansion automatically preserves 
all those sum rules and general theorems that can be expressed by the
action of linear operators (integrations, Fourier transforms, functional
derivatives and any other) over the generating functional (see, e.g.,
\cite{Ce-90}).

On the other side, a system of nucleons and pions seemed to us a good 
laboratory to study how a renormalizable quantum-field-theoretical model 
embeds into a many-body problem.

\begin{figure}
\begin{center}
\mbox{
\epsfig{file=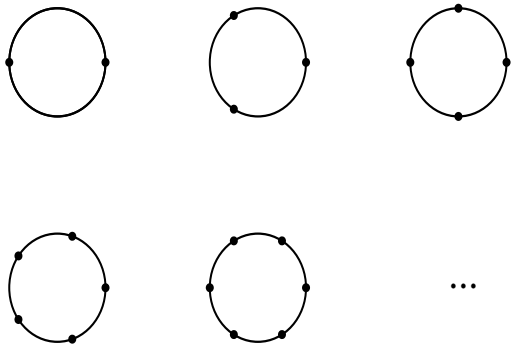,height=6cm,width=8cm}
}
\end{center}
\caption{\protect\label{fig1} The bosonic effective action. Dots denote
the external points $x_i$ while solid lines stand for in medium free nucleon
propagators.}
\end{figure}

The idea of bosonization arose quite naturally there,
as it was obtained by representing the 
generating functional of the system by means of a Feynman path integral 
and by explicitly integrating over the fermionic degrees of freedom. 
Then the system turned out to be described by the bosonic 
effective action
\begin{eqnarray}
\lefteqn{S_B^{\rm eff}=\frac{1}{2}\int 
d^4x\,d^4y\,\phi(x)\left[D_0\right]^{-1}(x-y)
\phi(y)}
\label{lt5}\\
&&-\sum_{n=2}^\infty\int d^4x_1\dots d^4x_n
\frac{1}{n}\Pi^{(n)}(x_1,\dots,x_n)\phi(x_1)\dots \phi(x_n)\;,
\nonumber
\end{eqnarray}
$\phi$ denoting the pion field (isospin is neglected 
for simplicity), $D_0$ its free propagator and the highly non-local vertices 
$\Pi^{(n)}(x_1,\dots,x_n)$, that keep memory of the fermion dynamics, being
shown diagrammatically in fig.~\ref{fig1}. 

This action generates a new class of approximations -- or a recipe to 
collect classes of Feynman diagrams together -- when the semiclassical 
expansion is carried out. This scheme is also referred to as boson loop 
expansion (BLE). Its renormalizability has been proved in 
\cite{AlCeMoSa-88}.

The practical recipe to classify a Feynman diagram according to its order 
in BLE is to shrink to a point all its fermion lines 
and to count the number of boson loops left out.

Thus the mean field level of the theory just coincides with the RPA (or,
better, with the ring approximation, without antisymmetrization).
This can be seen either diagrammatically, according to the above
rule, or observing that the quadratic part of the action just contains 
the inverse ring propagator, namely
$\frac{1}{2}D_0^{-1}-\frac{1}{2}\Pi^{(2)}\;.$
Note that $\Pi^{(2)}$ coincides, up to factors, with the Lindhard 
function. The present unusual notation is chosen 
to agree with refs. \cite{CeSa-88,CeCoCoSa-92}.

\begin{figure}
\begin{center}
\mbox{
\epsfig{file=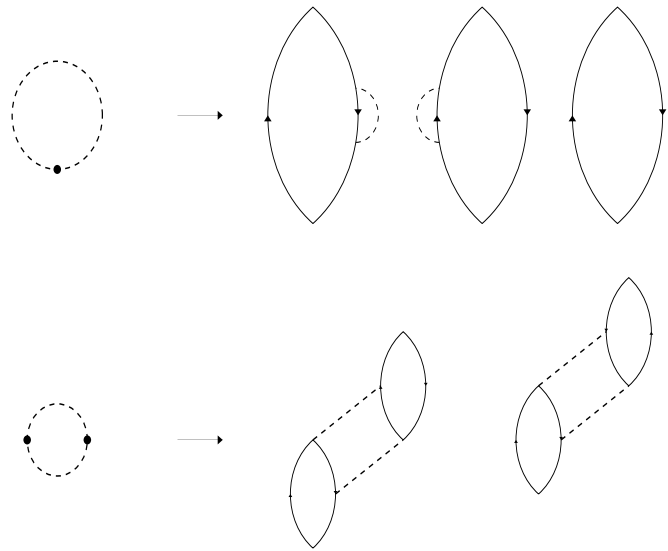,height=8cm,width=10cm}
}
\end{center}
\caption{\protect\label{fig2} First order diagrams in the BLE. Recall that the
dashed lines always describe RPA-dressed bosons}
\end{figure}
At the linear response level the mean field 
is thus described by the RPA-dressed polarization propagator if the 
probe has the same quantum numbers of the pion or, if not, by the bare Lindhard 
function.

At the one-loop order the only possible diagrams (we neglect here the 
MEC, since we shall not be concerned with them in the rest of the paper) 
are those of fig. \ref{fig2}. 
We remind once more that all bosonic lines have 
to be seen as RPA-dressed.

\begin{figure}
\begin{center}
\mbox{
\epsfig{file=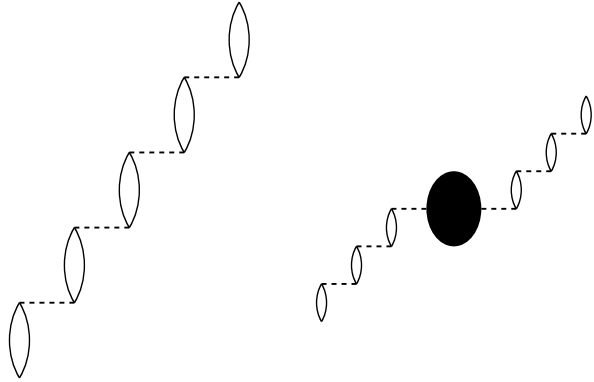,height=6cm,width=9cm}
}
\end{center}
\caption{\protect\label{fig2bis} Diagrams pertaining to the response 
function at the one-loop level. The black bubble summarizes the 5 
diagrams of fig.\protect\ref{fig2}; we explicitly indicated the RPA
dressing of the bare meson propagators (dashed lines).}
\end{figure}

To be more specific, the full response at the one-loop order is given 
(up to factors) by the imaginary part of the diagrams of fig. 
\ref{fig2bis}, where the black bubble denotes the sum of all the 
diagrams of fig. \ref{fig2}.

It is evident from fig. \ref{fig2} that only fermion loops with at most four 
external legs intervene at the one-boson-loop order. This fact is of 
overwhelming relevance since analytical expressions are available for them 
\cite{CeSa-88,CeCoCoSa-92}, at least in the nonrelativistic kinematic.

This occurrence makes the calculation feasible and in ref. 
\cite{AlCeMoSa-90} we got some preliminary results for the longitudinal 
response on the QEP, the only dynamics being there the pion exchange.
A very tricky technicality forced us to consider static pions only. 
We are not able at present to perform a relativistic calculation because the 
outcomes of \cite{CeSa-88,CeCoCoSa-92} cannot be fully translated to 
a relativistic frame.
It is well known that the pion branch in the RPA scheme is coupled to the 
particle-hole mode. Within a nonrelativistic nucleon kinematics, 
in a momentum range around 1.6 GeV/$c$, the pion branch enters 
the particle-hole (p-h) continuum, because the former is relativistic and the 
latter is not. This originates wild oscillations which 
are by one side absolutely unphysical, but on the other numerically
unavoidable. Thus we were forced to accept a static version of the pion 
exchange potential -- at least for the moment:
some promising work is in progress to overcome this limitation.

Further, the pion condensation is unavoidably met in the RPA scheme
if we only allow pion exchange without accounting for short range correlations
(SRC). Since pion condensation, if any, occurs very far from 
nuclear matter stability region, we are again forced to forbid its occurrence 
by phenomenologically embedding SRC in our model by means of the
Landau parameter $g'$, which resumes all the complicated short range 
p-h interactions in the pion channel.

To adhere to phenomenology we are thus forced to change somehow our 
approach, both in the formalism and the underlying philosophy.

Concerning the former we must give up a field theory 
and adopt a potential frame. The derivation of the response at the one 
loop level is given in detail in ref. \cite{CeSa-94} and will not be 
repeated here. We only remind the reader that a Hubbard-Stratonovitch 
transformation \cite{MoGaNaRa-74,Ke-70,Kl-78} bosonizes a potential theory, 
leading to an effective action coincident with (\ref{lt5}) up to a 
redefinition of the symbols: $D_0^{-1}$ is replaced by the inverse 
potential in the given p-h channel and the field $\phi$ is
reinterpreted as an auxiliary field. The topology of the diagrams 
remains unchanged.

To exemplify we used in \cite{AlCeMoSa-90} the effective interaction
\begin{equation}
D_0(q)\longrightarrow
V_\pi(q)={f^2_{\pi NN}\over m_\pi^2}\left\{g^\prime_L({\bf q})-
{{\bf q}^2\over m^2_\pi+{\bf q}^2}\right\}v_\pi^2(q^2)\;,
\label{lt6}
\end{equation}
where the function $g^\prime_L({\bf q})$ embodies the SRC.

Leaving aside technicalities (discussed, on the other 
hand, in \cite{CeSa-94,AlCeMoSa-90}) eq. (\ref{lt6}) deserves some 
conceptual comments. Not only in fact we were forced to abandon a 
field-theoretical model, but the parametrization of SRC also 
implies the replacement of a realistic potential with an effective 
one, ruled by the Landau parameter $g^\prime_L$. 

The usual scheme where the 
Landau-Migdal parameters are used is, in our language, the mean 
field level; this approximation underlies also the phenomenological 
analyses used to substantiate them (see for 
instance \cite{SpWeWi-77}). We are working now at a deeper level, namely 
the one-boson-loop one. Here the need 
of the Landau-Migdal parameters still survives, but they cannot 
be directly derived from the same phenomenology.
The occurrence in a diagram of two subsequent Landau parameters 
renormalizes their value. Thus we are not linked 
to preserve the same values as at the mean field level
but we should look to them as to free 
parameters able to reproduce the experimental data. 

In practice we attempted, as far as possible, to maintain some 
connection with the known phenomenological analyses. This should be mandatory
in a purely nucleonic scheme because if the loop expansion is convergent, 
then the parameters
redetermined at the 1-loop order should be not so different from
their corresponding mean field values. The presence of intermediate 
states with one or more nucleons excited to $\Delta$'s changes 
qualitatively the previous statement, as is the case for the so-called
``box diagrams''. We will discuss better this point
in the following.
Thus ultimately we introduced as phenomenological ingredients of our 
models the effective interactions in the various channels
to reproduce reasonably well the longitudinal and transverse response 
functions, nevertheless trying to preserve some adherence to the Landau 
parameters as fixed at the mean field level when no other indications 
are available.

However, in going from the mean field to the one loop level, a 
qualitative change is met. In fact in the former case 
not too high momenta (i.e., of the order of
the  momentum of the probe) enter the dynamics, while in the latter
the bosonic loop momentum is integrated over and the high momentum behaviour 
of the effective interaction rules the convergence of the integral (see,
e.g., \cite{Ce-93} for a more detailed discussion).
This is exploited in 
(\ref{lt6}) by forcing a $q$-dependence in 
$g^\prime_L$ such that 
\begin{equation}
g^\prime_L(q)\stackrel{q\to\infty}{\longrightarrow}1\;,
\label{lt7}
\end{equation}
in order to account for SRC. But this in turn requires the introduction of 
a further cut-off related to the SRC range that becomes the crucial 
parameter ruling the one loop corrections. 

These last considerations set the philosophy underlying our approach.
To resume, the effective interaction is a phenomenological input, to 
be fixed, as far as possible, on the experimental data.
Two quantities characterize the effective interaction, namely 
$g^\prime$ (taken at $q\to0$), which mainly governs the response at the 
mean field level, and the cut-off which sets the behaviour of the 
effective interaction at high momenta, which instead determines the size 
of the one-loop corrections.

So far we have only  discussed a system of nucleons and only one 
effective interaction. Needless to say, the dynamics required to describe
the nuclear responses is by far richer. Formally our approach remains 
unchanged and other channels are accounted for by simply interpreting the 
dashed lines in figs. \ref{fig2} as a sum over all the allowed channels,
as we shall discuss in the next section.

Also, the excitation of the nucleons to a resonance (in the present 
paper we will be concerned with the $\Delta_{33}$ only) can be 
allowed by interpreting each solid line as a nucleon or as a $\Delta$ 
and again summing over all possible cases. {\em Remarkably the topology 
of the diagrams does never change}. Simply, conservation laws
will kill some of them.
\section{The dynamical model \label{sect4}}

To substantiate the theoretical frame discussed in the previous section, 
we need to define the dynamics we are going to consider. 
In this respect, three topics,
namely the effective interactions, the diagrammology -- with particular 
attention to the excitation of a $\Delta$-resonance -- and the external 
current characterize the model.

In the above we confined ourselves, just to simplify the 
exposition, to a very poor scheme, because the 
extension to a richer dynamics is formally trivial.
The same cannot be said however about the physical content. 

In ref. \cite{AlCeMoSa-90} we 
evaluated the longitudinal response using the effective 
interaction (\ref{lt6}) in the pion channel. The results were 
qualitatively unsatisfactory, but they went in the right direction.
A more sophisticated approach was pursued in \cite{CeSa-94}: the 
formalism is the same, but other channels are considered.
We focused there our attention on the pion channel, the $\rho$ channel 
and the transverse $\omega$ channel. We neglected the scalar channels 
following the indications of Speth et al. \cite{SpWeWi-77} that suggest 
there a weak or vanishing effective interaction. Of course this outcome 
is questionable and was adopted in \cite{CeSa-94} only for sake of
simplicity. The statement of ref. \cite{SpWeWi-77} refers in fact to the
mean field level, while at the one loop order effective interactions may
exist able to lead, owing to cancellations between them and to the interference
with the mean field level, to the same phenomenological outcome
of ref. \cite{SpWeWi-77}. At the mean field level, in fact, the statement 
that the effective interaction in a given spin--isospin channel is small 
means that the linear response in the chosen channel is close 
to the one of the FFG.
At the one loop level instead 
the effective 
interaction in a given channel can affect the responses in all the other 
ones.
Thus we
must build up a coherent set of effective interactions in all the 
channels such that, for instance, in the scalar channels the global effect of 
the mean field and of the one loop corrections  cancel together, being 
however both not necessarily small.
Further, the excitation of a nucleon to a $\Delta$-resonance was only 
partially accounted for (and largely underestimated)
in \cite{CeSa-94}. It will play instead a 
central role in the present paper. 

We shall consider here the (correlated) exchange of $\pi$, 
$\rho$ and $\omega$-mesons.
In each particle-hole channel the interaction will read
\begin{equation}
V_m(q)={f^2_{\pi BB^\prime}\over m_\pi^2}\left\{g^\prime_m({\bf q})-
C_m{{\bf q}^2\over m^2_m+{\bf q}^2}\right\}v_m^2(q^2)\;,
\label{lt3}
\end{equation}
where the index $m$ runs over the accounted channels and $B=N$ or 
$\Delta$.

Concerning the pion, it propagates in the isovector spin-longitudinal 
channel. Since the pion exchange is 
not far from being perturbative, its coupling constant is not assumed to 
be renormalized in the medium and hence, by definition, $C_\pi=1$. 

A vector meson interact with the nucleons by means of the relativistic 
lagrangian
\begin{equation}
g\overline{\psi}\gamma_\mu\psi\phi^\mu+\frac{f}{4m}\overline{\psi}
\sigma_{\mu\nu}\psi(\partial^\mu\phi^\nu-\partial^\nu\phi^\mu)
\label{vmrl}
\end{equation}
where the first term is a standard coupling to a current while the
second term describes an ``anomalous magnetic moment''.
According to the analyses of ref. \cite{MaHoEl-87} this last term is 
absent in the $\omega$ case but dominates the $\rho$-exchange.

Obviously the interaction (\ref{vmrl}) generates a convective and a spin 
current coupled to the 3-vector part of the mesonic field
plus a scalar coupling to its time component. Customarily the convective 
current is neglected. Thus we are left with an interaction in the 
spin-transverse channel coming from the spin current and one in the 
scalar channel coming from the coupling to $\phi^0$. Since the 
``anomalous magnetic moment'' is absent in the $\omega$ case, the 
coupling constants of these two interactions would coincide, while
the scalar component of the $\rho$ is coupled much more weakly than its 
3-vector part.

Coming now to the effective interactions to be put in our machinery, we 
adopted for the $\rho$ case the same coupling constants as in the Bonn 
potential \cite{MaHoEl-87}, corresponding to a $C_\rho\sim  2.3$ in the 
isovector spin-transverse channel and to $C_\rho\sim 0.05$ in the scalar 
isovector one, and this because, in spite of its large coupling the 
$\rho$-exchange is perturbative too due to the high mass of the 
$\rho$ itself. Further, in the spin-transverse channel we need to introduce 
SRC, because in the Landau limit the effective interaction must equal 
the one of the spin-longitudinal channel, and this makes by far 
non-perturbative the whole effective interaction.
We shall see later that the particular form of this interaction plays a 
crucial role in explaining the responses. We shall also account for the 
interaction of the $\rho$ meson with the nuclear medium other than ph 
and $\Delta$h excitations (we means for instance $\rho\to\pi\pi$ with 
further interaction of pions with the medium) by attributing to the 
$\rho$ (in the spin-transverse channel only) an effective mass to be 
specified later. In the scalar isovector channel no Landau parameter has 
been added and the mass is left unchanged.

Coming to the $\omega$ mesons, we have left the interaction in 
the isoscalar spin-transverse channel unchanged, since it is 
weak, but the scalar isoscalar channel needs to be renormalized 
drastically, because there the effective interaction is made by the 
exchange of many $\omega$ mesons. Since we have no sufficiently striking 
experimental information on the details of this effective interaction we 
have simply taken the corresponding $C_{\omega}$ as a free parameter, 
without changing nor the functional form neither its range. In order to 
avoid uncontrolled complications no Landau parameters have been added
in these two channels.

The Landau parameter $g^\prime$ is thus present in the isovector
spin-transverse and spin longitudinal channels only, where
it has been parametrized according to
\begin{equation}
g^\prime_{L,T}(q)=C_m+(g_0^\prime-C_m)\left[{q_{c\,{L,T}}^2\over
q_{c\,{L,T}}^2+q^2}\right]^2
\label{lt8}
\end{equation}
in such a way that for $q\to 0$ $g^\prime_{L,T}(q)\to g^\prime_0$ and for 
$q\to\infty$ a proper generalization of condition (\ref{lt7})  is respected,
namely
\begin{equation}
g^\prime_{L,T}(q)\stackrel{q\to\infty}{\longrightarrow}C^{T=1}_{L,T}\;.
\label{lt7.1}
\end{equation}
The further 
cut-offs are called here $q_{c\,{L,T}}$. The form we assumed for the effective 
interaction is not qualitatively different from other widely employed 
in the literature (see for instance \cite{BrBaOsWe-77,OsSa-87}), the 
only true change being that the value at ${\bf q}=0$ is in our approach 
free. Up to our knowledge there is no experimental input to fix 
$q_{c\,L}$, but it is commonly assumed of the same order of the $\omega$ 
mass (and hence of the inverse radius of the repulsive core in the 
nucleon-nucleon interaction) and many calculations using this 
value have been performed successfully \cite{OsWe-79,OsWe-79a}.
No indications are available instead for $q_{c\,T}$. Some possible 
choices are discussed in \cite{CeCoSa-96}, but since the interaction 
in the $\rho$-channel dominates the responses, as we shall see below, we 
defer its description to the subsequent sections.

The vertex function present in (\ref{lt7}) is customarily 
assumed to be
\begin{equation}
v_m(q^2)={\Lambda_m^2-m_m^2\over \Lambda_m^2 + {\bf q^2} -q_0^2}
\label{lt9}
\end{equation}
but in our calculation has been assumed as static too for sake of 
coherence.

\begin{figure}
\begin{center}
\mbox{
\epsfig{file=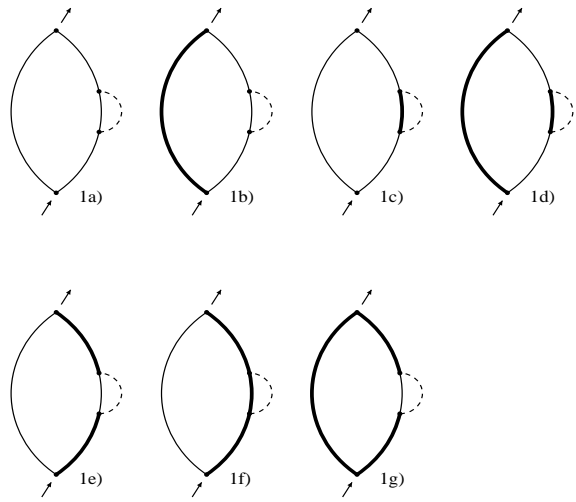,height=7.5cm,width=10cm}
}
\end{center}
\caption{\protect\label{fig3} The self-energy diagrams with inclusion of
$\Delta$-resonances. The $\Delta$'s are described by thick lines}
\end{figure}

\begin{figure}
\begin{center}
\mbox{
\epsfig{file=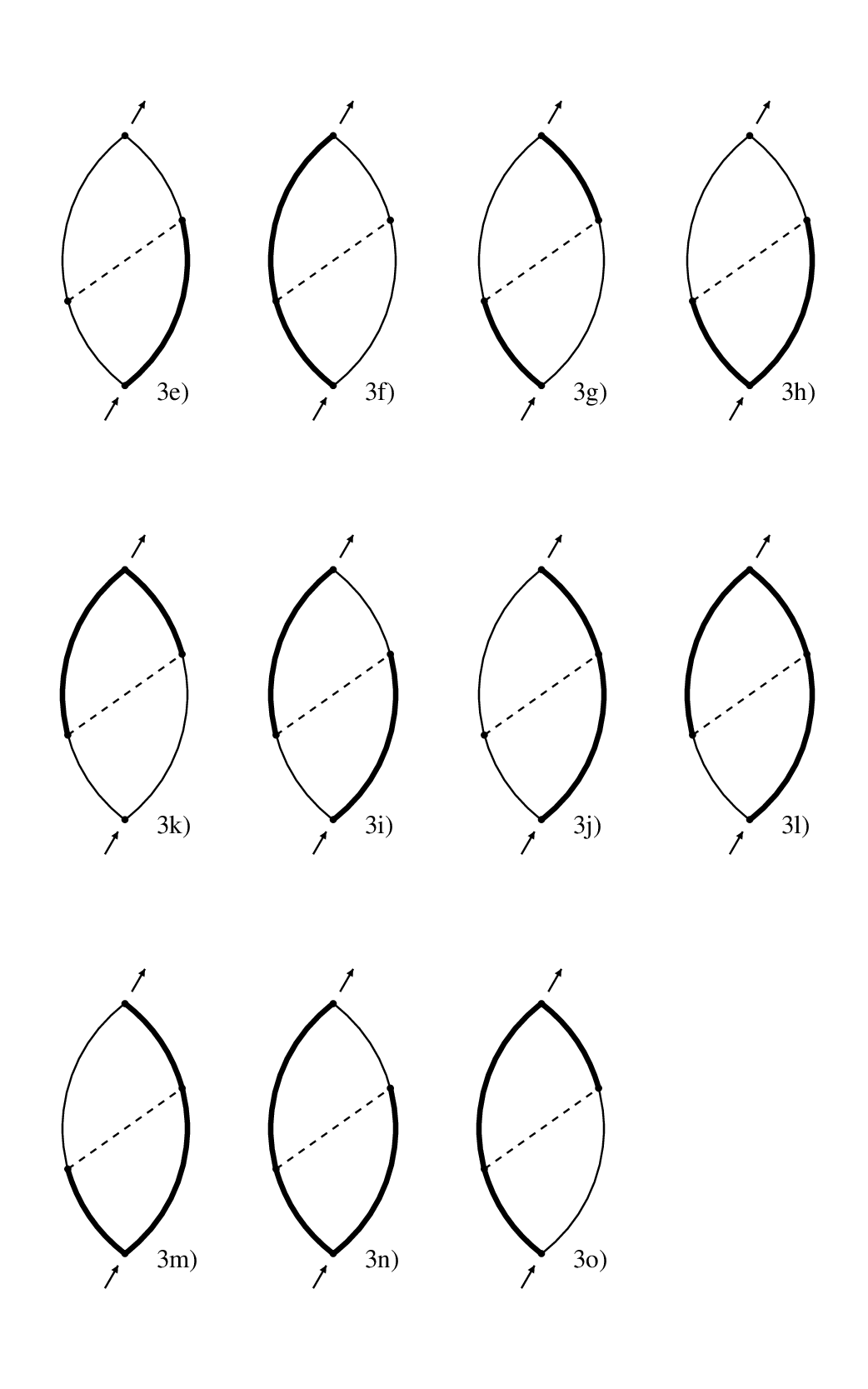,height=12cm,width=10cm}
}
\end{center}
\caption{\protect\label{fig4} The exchange diagrams with inclusion of
$\Delta$-resonances. The $\Delta$'s are described by thick lines}
\end{figure}

\begin{figure}
\begin{center}
\mbox{
\epsfig{file=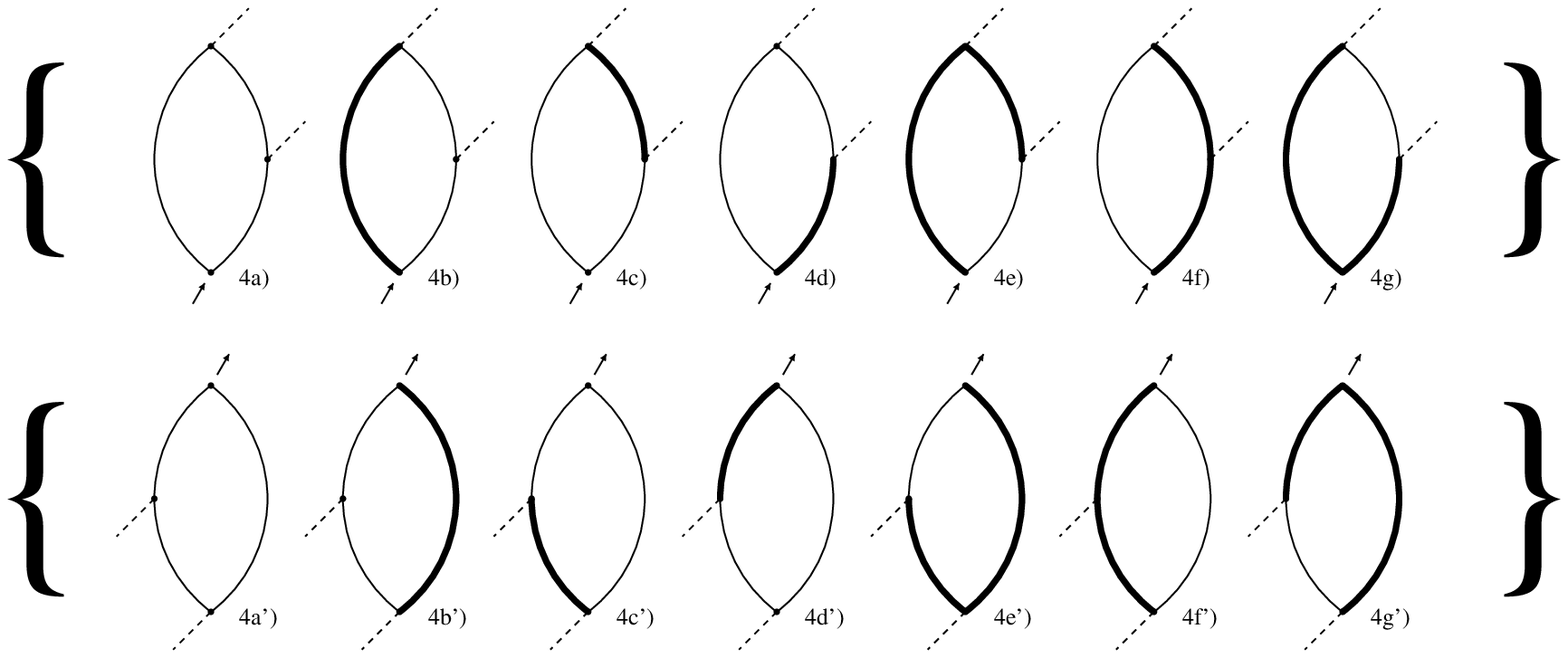,height=8cm,width=12cm}
}
\end{center}
\caption{\protect\label{fig5} The correlations diagrams with inclusion of
$\Delta$-resonances. The diagrams are obtained by combining each 
component of the upper line with each piece of the lower one.
The $\Delta$'s are described by thick lines}
\end{figure}

Next we come to the role of the $\Delta$-resonance. 
The diagrammology, as far as its topological content is concerned, 
follows quite simply from fig. \ref{fig2}, by allowing each line to be 
either a nucleon or a $\Delta$. Isospin conservation rule cuts some 
diagram but too many of them survive. They are displayed in figs. 
\ref{fig3} to \ref{fig5}. 

Fig. \ref{fig3} expands the first and second 
diagram of fig. \ref{fig2} (self-energy terms), fig. \ref{fig4} expand 
the third one (exchange) and fig. \ref{fig5} the last two diagrams
(correlation terms). Note that these 
are obtained by expanding both the two bubbles present there. Since each 
bubble translates into 7 possible diagrams, then 49 of them 
arise for each of the two terms accounted for. 

In figs. \ref{fig3} to \ref{fig5} only the isospin conservation law has 
been considered.

The above shown diagrammology deserves some comments concerning the
feasibility of the calculation. There are in fact a total of 125 
Feynman diagrams (14 self-energy, 15 exchange and 98 correlation 
diagrams). Since we consider five different effective interaction, any 
dashed line should be counted 5 times, giving thus a total a 2595 diagrams.
To understand the relevance of the results of refs. \cite{CeSa-88,CeCoCoSa-92}
in enabling us to truly evaluate all of them consider the following:
the usual way of computing them is to reduce each one to Goldstone 
diagrams by explicitly performing the frequency integration. In the 
present case the integration over the bosonic frequency cannot be simply 
performed, due to its RPA-dressing. Would it be even possible, due to the 
all the occurring time ordering that must be accounted for, one should 
end up with 1767480 (!) diagrams almost all containing 9-dimensional 
integrals with extremely complicated 
boundaries. In any case, a hopeless task.

Refs. \cite{CeSa-88,CeCoCoSa-92} tell us that the integrals over the 
fermionic loops can be performed analytically {\em provided we sum up 
all the Goldstone diagrams}, i.e., at the Feynman diagram level. This 
not only shortcuts the number of diagrams, coming back to the 2595 ones 
quoted before, but it reduces the numerical integrations to the 
bosonic loop only. Since here one integration is trivial we are left 
out in practice with 2595 3-dimensional integrals,
a tedious but not impossible job
(as an aside, further tricks have been of course used so to reduce both the 
computing time and the structure of the calculation by parallelizing the 
integration procedure and by exploiting the regularities of the 
diagrams, so to reduce the codes to a manageable form and the computing 
time to a reasonable order -- to say, one day for a complete calculation 
with one set of parameters on a last generation  $\alpha$-VAX). 

Still the Feynman rules have not been fully determined. 

First we need to 
define the $\Delta$ propagator, and this is done by setting
\begin{equation}
G_\Delta(p)=\frac{1}{p_0-\frac{{\bf p}^2}{2 M_\Delta}-\delta M+i\eta}
\label{lt10}
\end{equation}
($\delta M$ being the mass difference between $\Delta$ and nucleon), 
without accounting for the $\Delta$ width. This is correct for two 
reasons: first, the width is frequency and momentum dependent and it has 
of course a threshold at the pion mass in the $\Delta$ rest frame
and since the $\Delta$ lives almost always below this threshold, the 
width is coherently set to 0; second, in the BLE the $\Delta$ width comes out 
-- at least partially -- from a one-boson-loop correction, like, e.g., 
in the diagram 1e) of fig. \ref{fig3} and, if any, is already accounted 
for.

A second need in order to evaluate the Feynman diagrams is to define the 
vertices. They carry in fact some spin-isospin matrices: 
in the $N$-$N$ vertex a spin-longitudinal potential 
brings with it a factor ${\bf q}\cdot \bfm \sigma$ while a 
spin-transverse one has the factor ${\bf q}\times \bfm\sigma$ and of course 
an isovector potentials adds to them a factor $\tau_i$; in a 
$N$-$\Delta$ vertex the Pauli matrices $\bfm\sigma$ and $\tau_i$ 
are replaced by the corresponding transition matrices $\bf S$ 
and $T_i$ \cite{BrWe-75} and finally  in the case of the
$\Delta$-$\Delta$ vertex we need the less known 3/2--3/2 spin 
matrices, denoted here by $\bfm{\cal S}$ and ${\cal T}_i$. They are 
displayed in appendix \ref{appA}. 

As a matter of fact we do not need their 
explicit expression, but only the trace of at most four of them. 
The traces needed to evaluate the spin-isospin part of our diagrams are 
also given in the appendix \ref{appA}.
The traces have to be evaluated for each of the 2595
Feynman diagrams and for the longitudinal and transverse isoscalar and 
isovector responses: actually the job is not so tedious as it could 
seem, owing to the many symmetries in the diagrams: the main ingredients 
needed to construct them are given in appendix \ref{appB}. These factors
have been anyway checked out for maximum safety with an analytical
calculation performed with the Mathematica code.

The last topic we need to discuss in order to complete the description of 
the model is the 
form of the e.m. current. It acts on three different sectors, namely the 
$N$-$N$, the $N$-$\Delta$ and the $\Delta$-$\Delta$ ones and can be 
separated into the longitudinal and transverse channel and, further, in 
its isoscalar and isovector components.

In the longitudinal channel, the simplest one, we account for the $N$-$N$ 
and $\Delta$-$\Delta$ charge operators but of corse {\em not} for 
$N$-$\Delta$ transitions. We have chosen the following form for the 
matrix elements of the charge operators:
\begin{eqnarray}
<N|\rho|N>&=&\frac{1}{2}\tilde G_E^{T=0}+\frac{1}{2}\tau_3\tilde 
G_E^{T=1}\\
<\Delta|\rho|\Delta>&=&\frac{1}{2}\left(1+{\cal T}_3\right)\tilde 
G_E^\Delta\;.
\end{eqnarray}
Here the $\tilde G_E$ are defined as 
\begin{equation}
\tilde G_E=\frac{G_E}{\sqrt{1+\tau}}
\label{lt12}
\end{equation}
to account for the recoil, according to \cite{De-84}. In the
nucleon case the familiar Sachs form factors 
are employed, while in the $\Delta$ case 
no experimental evidence can be invoked to select between different 
theoretical models. We have chosen here for $G_E^\Delta$, as a 
reasonable compromise, the dipole 
form, which should better account for the high momentum tail, with the 
radius coming from the constituent quark model
\cite{Gi-90}, where it comes out to be a 27\% 
higher than the one of the nucleon. We correspondingly increase the 
cut-off in the dipole form factor of the same amount.

The current matrix elements instead has a more complicated structure. In 
the nucleonic sector we are forced to the shortcut of dropping the 
convective part of the current by putting
\begin{equation}
<N|{\bf j}|N>=i\frac{{\bfm \sigma}\times{\bf q}}{2m}G_M \;.
\label{lt15}
\end{equation}

The transition current for the single nucleon is written very simply in 
the centre-of-mass frame. We retain the $M1$ component only, so that the 
current reads
\begin{equation}
<N|{\bf j}|\Delta>\bigm|_{\rm c. of. m.}
=i\mu_{N\Delta}T_3\frac{{\bf S}\cdot {\bf q}\times 
{\bfm\epsilon}}{2m}G_M^\Delta\;;
\label{lt16}
\end{equation}
we need however in our calculations, to be able to use the results of 
refs. \cite{CeSa-88,CeCoCoSa-92}, an equivalent expression in the 
laboratory frame. The transformation of (\ref{lt16}) to the laboratory 
frame generates a convection part, which is forcedly neglected, and, to 
the lowest order in the non-relativistic expansion, the replacement
\begin{equation}
{\bf q}\to \frac{m}{m+\omega}{\bf q}\;.
\label{lt20}
\end{equation}

Finally the current in the $\Delta$ sector is written as
\begin{equation}
<\Delta|{\bf j}|\Delta>=i\mu_\Delta\frac{1+{\cal 
T}_3}{2}\frac{{\bfm{\cal S}}\times{\bf q}}{2m}G_M^\Delta
\label{lt17}
\end{equation}
where $\mu_\Delta=2.95$ according to \cite{Gi-90} and again, both in 
(\ref{lt16}) and (\ref{lt17}) we have 
taken $G_M^\Delta$ in dipole form with the same radius as $G_E^\Delta$.

A relevant issue of the present work is that the transition current is 
present in the magnetic channel but not in the charge-longitudinal one. This 
seemingly trivial consideration implies the highly nontrivial 
consequence that many of the displayed diagrams are killed in the 
charge-longitudinal channel. More explicitly only the diagrams 1a), 1c) 
and 1g) survive in fig. \ref{fig3} (and the same happens to the 
second self-energy diagram of fig. \ref{fig2}), in the exchange diagram 
only 3a), 3h), 3k) remain and finally in the correlation diagrams only 
the combinations of 4a), 4c) and 4g) with 4a'), 4c') and 4g') are 
allowed. This means that 21 diagrams (including self-energy and 
exchange) survive in the longitudinal channel 
while all 127 diagrams survive in the transverse one. 

Thus we realize in practice what argued in ref. \cite{ErRo-86}, where it 
was outlined that the so called ``swollen nucleon'' hypothesis, which is 
channel-insensitive in a bag model, becomes channel-dependent when the 
mesonic cloud is accounted for. Here the channel dependence comes out 
at the one-loop level and is represented by the many diagrams 
allowed only in the transverse response. They individually could well 
contribute by a very 
small amount, but their large number provides at the end of the 
calculation a non-negligible result.

\section{The mean field level \label{sect5}}

The mean field level in a coherent discussion is expected to precede the 
first order corrections, and we shall follow this aptitude in the present 
exposition. The situation at hand is particularly unlucky however,
because the mean field level is inextricably linked in the present case 
to the one-loop corrections. In fact:
\begin{enumerate}
\item
In the charge-longitudinal channel the mean field is mainly dominated by 
the $\omega$ exchange, which is repulsive and so entails a quenching of 
the response. As a simple mathematics shows, this implies also a 
hardening of the response \cite{AlCeMo-89} not observed in the 
data. Further, we neglect the $\sigma$-meson exchange, 
which is known from the Dirac phenomenology \cite{Wa-74,ClHaMe-82,Cl-86}
to be attractive and that cancels to a large amount with the $\omega$ 
exchange. In our approach, in the same line of thought of the Bonn 
potential,\cite{MaHoEl-87,Ho-81} the $\sigma$-meson is replaced 
by the box diagrams, i.e., two meson exchange with simultaneous 
excitation of one or two nucleons to a $\Delta$ resonance. These 
diagrams are contained indeed in our correlation diagrams,
which are however at the one-loop level.
\item
In the following section we will show that in the transverse channel the 
one-loop corrections are strongly suppressed (we shall see there that the 
$\Delta$-excitation makes this job). Thus the mean field becomes dominant in 
explaining the response and it is conversely ruled by the effective interaction 
in the $\rho$ channel. Thus, owing to this outcome, we can directly extract 
information about the effective interaction from the data, at least in this
spin--isospin channel.
\end{enumerate}

Let us now shortly review the present phenomenology and compare it with a 
FFG calculation. 

\begin{figure}
\begin{center}
\mbox{
\begin{tabular}{cc}
\epsfig{file=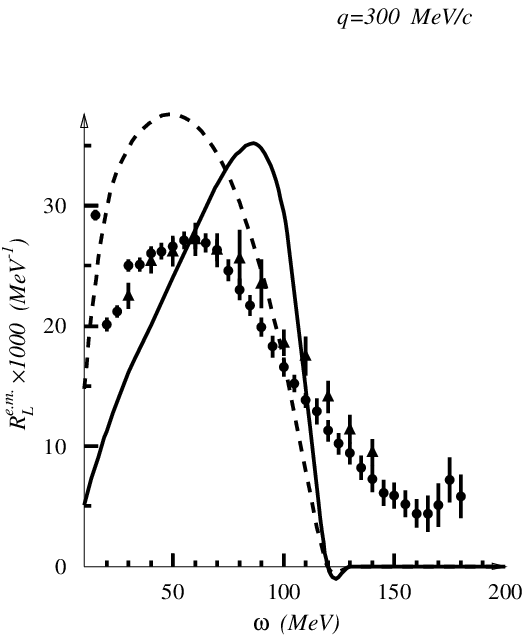,height=7cm}
&
\epsfig{file=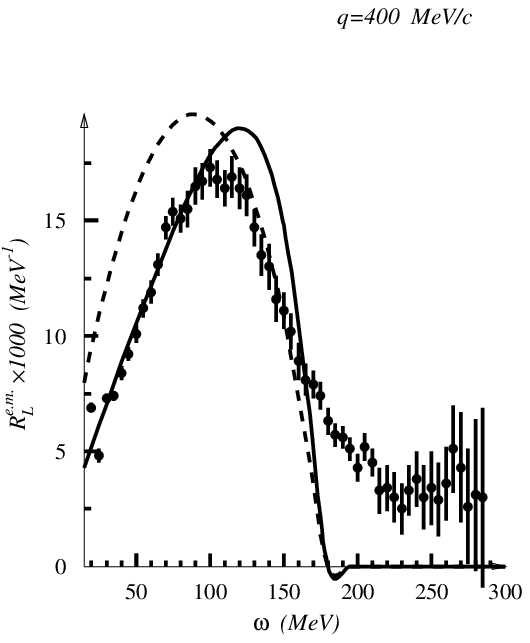,height=7cm}
\cr
\epsfig{file=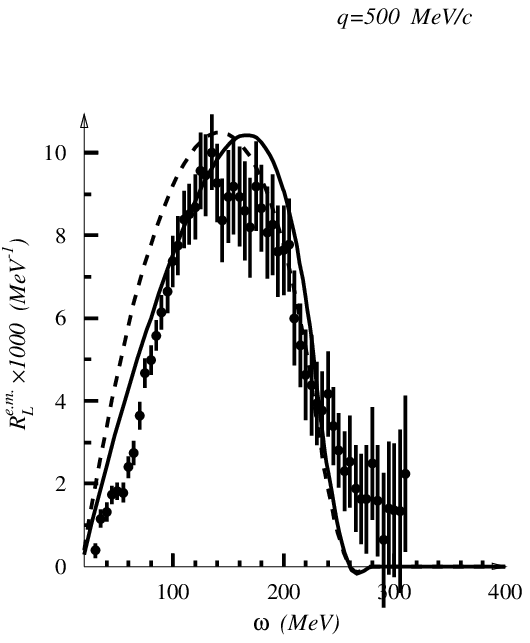,height=7cm}
&
\epsfig{file=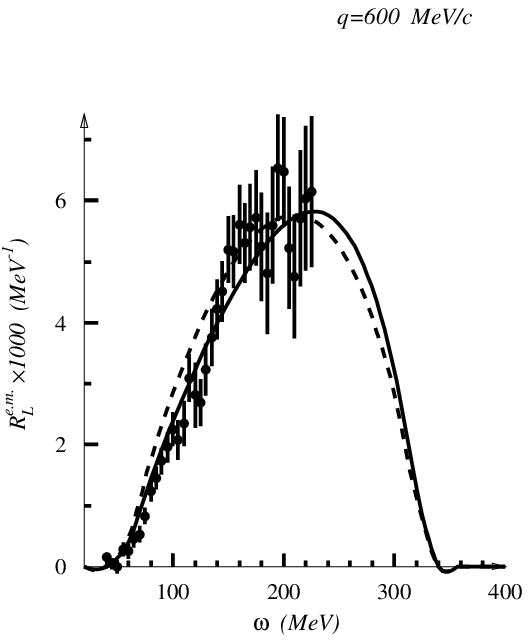,height=7cm}
\end{tabular}
}
\end{center}
\caption{\protect\label{fig6} Mean field calculation for 
the charge longitudinal response on $^{12}C$ at different transferred momenta.
Solid line: mean field; dashed line: Free Fermi Gas.
Data from \protect\cite{Me-al-84,Me-al-85} (circles) and from 
\protect\cite{Jo-95,Jo-96} (triangles). In the calculations we assumed 
$\k=1.10$ fm$^{-1}$}
\end{figure}

\begin{figure}
\begin{center}
\mbox{
\begin{tabular}{cc}
\epsfig{file=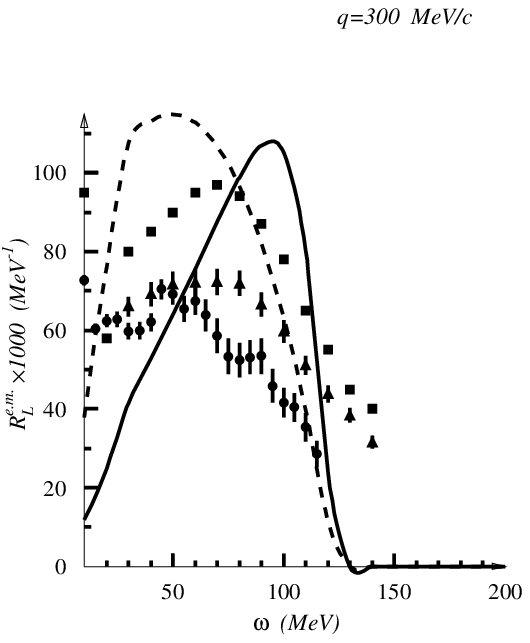,height=7cm}
&
\epsfig{file=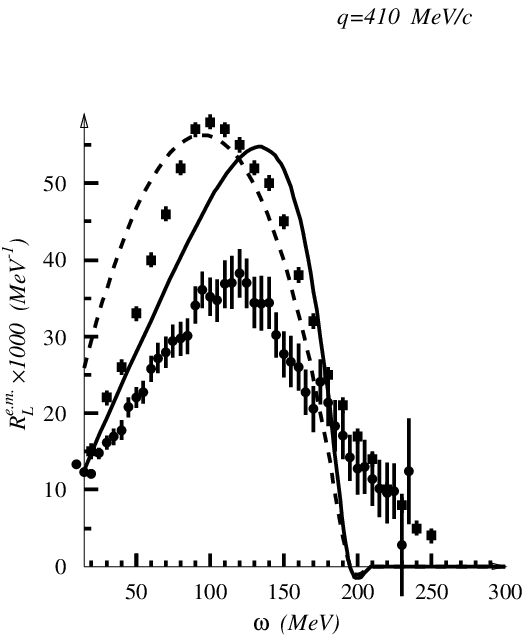,height=7cm}
\cr
\epsfig{file=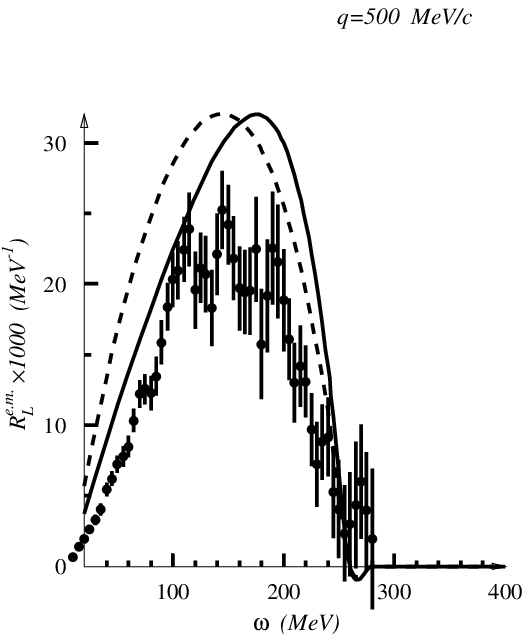,height=7cm}
&
\epsfig{file=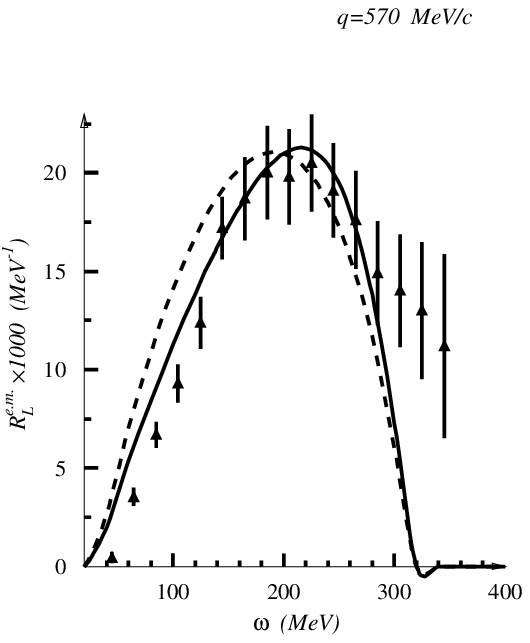,height=7cm}
\end{tabular}
}
\end{center}
\caption{\protect\label{fig7} 
Mean field calculation for 
the charge longitudinal response on $^{40}Ca$ at different transferred momenta.
Solid line: mean field; dashed line: Free Fermi Gas.
Data from \protect\cite{Me-al-84,Me-al-85} (circles), from 
\protect\cite{Jo-95,Jo-96} (triangles) 
 and from \protect\cite{Ya-al-93} (boxes).
 In the calculations we assumed $\k=1.2$ fm$^{-1}$
}
\end{figure}

We displayed in fig. \ref{fig6} the mean field previsions for the charge 
longitudinal response for 
$^{12}C$ at different transferred momenta and the same is done for 
$^{40}Ca$ in fig. \ref{fig7}

A common trend comes out clearly, namely that the FFG
is not able to explain the responses at low $q$.
The RPA dressing is carried out with $C_\omega$=0.15 and $C_\rho$=0.05:
its effect is clearly weak, but induces a remarkable hardening of the 
response. 
At higher $q$ instead the FFG becomes more and more suitable. 
We interpret this outcome as an evidence that correlations 
(both short-ranged and RPA-induced) are fading away, in agreement 
with our theoretical previsions -- again, as we shall see in sect. 
\ref{sect6}.

The situation of $^{40}Ca$ deserves a further comment. The FFG
fails to reproduce the data at $500 MeV/c$, those data being taken at 
Saclay. Remarkably in the Rosenbluth separation of 
Jourdan\cite{Jo-96} there is a qualitative agreement with the results of 
Saclay at $300 MeV/c$ while at higher momentum transfer the longitudinal 
response significant discrepancies arise. And in fact the Jourdan data
on $^{40}Ca$ at $570 MeV/c$ are in good agreement with the  free Fermi 
gas model, thus following the same trend of $^{12}C$.

The situation is drastically different in the transverse channel. There 
the effective interaction  is carried 
by $\rho$-meson exchange (attractive) plus SRC (repulsive). It is evident 
from figs. \ref{fig9} and \ref{fig10} that the FFG fails 
badly even at high momentum transfer. Here we do not expect a big 
contribution from the MEC (that part of them, at least, which is not yet 
embodied in our model, like, e.g., direct $\Delta$ excitation) 
at least in the left side of the QEP. This onset 
is supported by the occurrence of the $y$-scaling in this region.
The pion-in-flight contribution is in fact associated to the breaking of 
$y$-scaling, and experiments show its occurrence only in the right side 
of the QEP (see for instance the review \cite{DaMcDoSi-90}). Thus the 
deformation of one half of the peak has to be attributed either to the 
effective interaction in the channel of the $\rho$ or to one-loop 
corrections. Let us assume, for the moment without proof, that the former 
case occurs. 

If so, the transverse response has the form
\begin{equation}
R_T=-\frac{1}{\pi}\Im\frac{\Pi_T^{T=1}}{1-V_{\rm eff}^{T=1}\Pi_T^{T=1}}
\label{lt21}
\end{equation}
(we have neglected here the $T=0$ for sake of simplicity, since it 
represent at most a 3\% of the total; in our calculations however 
this component is also accounted for). $\Pi_T$ denotes the 
transverse component of 
\begin{equation}
\Pi_{\mu\nu}(x,y)=<\Psi_0|T\left\{j_\mu(x),j_\nu(y)\right\}|\Psi_0>
\label{pimunu}
\end{equation}
and $V_{\rm eff}^{T=1}$ is the 
effective interaction in the $\rho$-channel. It is well known 
\cite{AlCeMo-89} that an attractive effective interaction enhances the 
response, while a repulsive one depletes it. Experimental data shows, in 
the case of $^{12}C$, that for $q=300~ MeV/c$ the FFG is 
roughly sufficient to describe the data, while at higher values it 
systematically underextimate them. Thus in order to meet the 
experimental situation an effective interaction is required such to vanish 
around $q=300~ MeV/c$ and to be attractive in higher $q$-regions.

In the form we have chosen the attractive part of the effective 
interaction comes from the true $\rho$-exchange and the repulsive one 
from the SRC. We must keep in mind however that this is 
simply a parametrization for the effective interaction, which is instead 
an unknown as a whole. To maintain the form (\ref{lt3}) and reasonable 
values for $g^\prime_T$ and $C_\rho$ we have slightly changed the 
$\rho$-mass, attributing to it an effective value of $600 ~ MeV$. This by
one hand is not unreasonable, as the $\rho$ itself, either 
directly or when it lives as a pion pair, feels the effect of the nuclear 
medium, which consequently renormalizes is mass.
On the other hand, however, we want to stress that unusual parameters
in (\ref{lt3}) could look badly, but they do not disturb too much provided
the whole effective interaction looks reasonable.

The parameters which seem to better describe the transverse response 
are:
$m_\rho=600 ~ MeV$, $g^\prime_0=0.35$, $q_{c,T}=1.3~GeV/c$, 
$C_\rho=2.3$, $\Lambda = 1.75~GeV/c$
(making here the universal choice, i.e., keeping all the $C_\rho$ 
equal). With this choice the tail of the interaction is quickly 
decreasing so that the expected enhancement of the response at high 
transferred momenta is missing, but a milder behaviour of the tail would 
overemphasize the one loop corrections.

\begin{figure}
\begin{center}
\mbox{
\epsfig{file=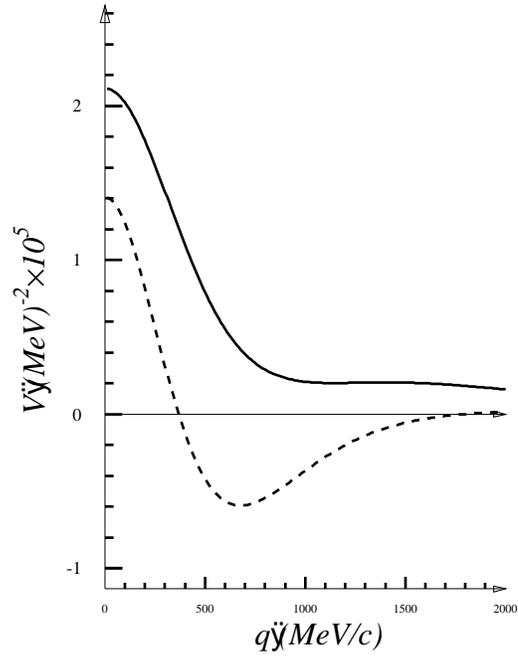}
}
\end{center}
\caption{\protect\label{fig8} 
The effective interaction in the isovector spin-transverse channel. 
Parameters of the solid line:
$m_\rho=770 ~ MeV$, $g^\prime_0=0.5$, 
$q_{c,T}=1.4~GeV/c$, $C_\rho=2.2$, $\Lambda = 2.5~GeV/c$;
parameters of the dashed line:
$m_\rho=600 ~ MeV$, $g^\prime_0=0.35$, $q_{c,T}=1.4~GeV/c$, 
$C_\rho=2.3$, $\Lambda = 2~GeV/c$.
}
\end{figure}

\begin{figure}
\begin{center}
\mbox{
\begin{tabular}{cc}
\epsfig{file=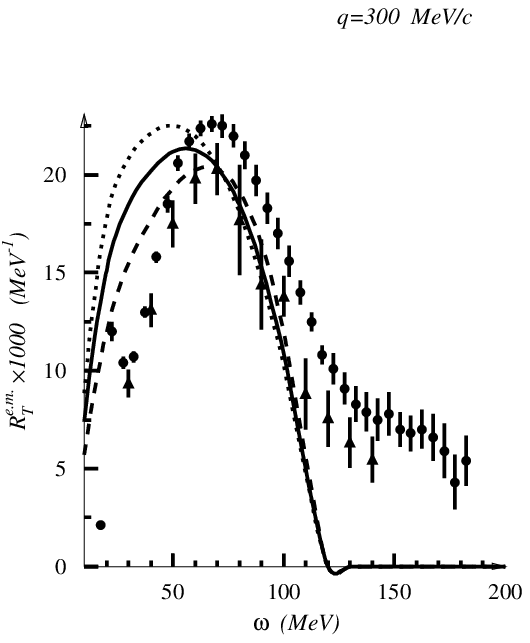,height=7cm}
&
\epsfig{file=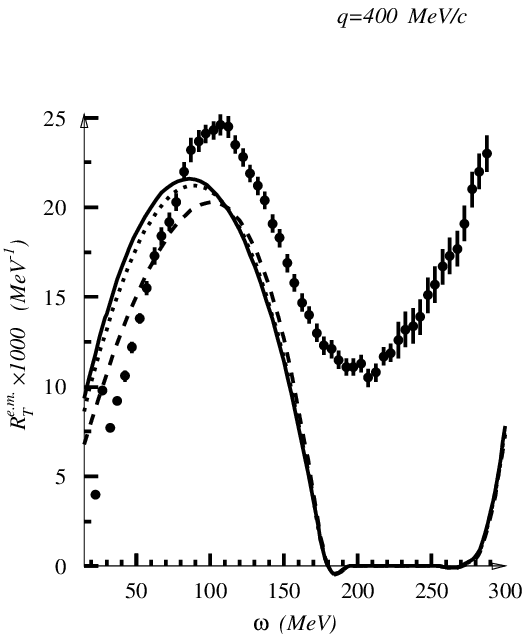,height=7cm}
\cr
\epsfig{file=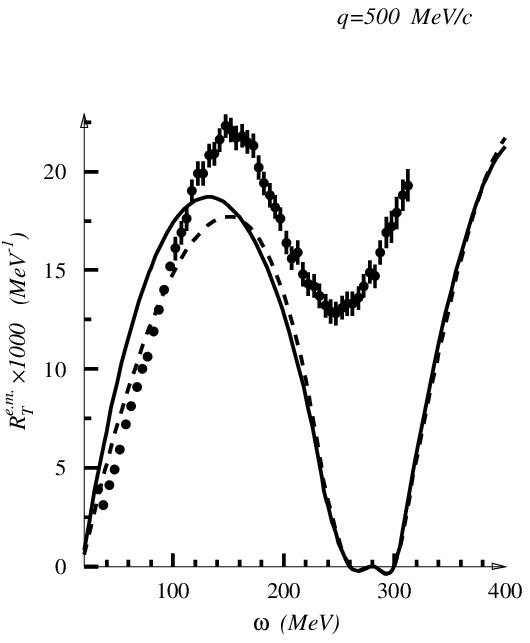,height=7cm}
&
\epsfig{file=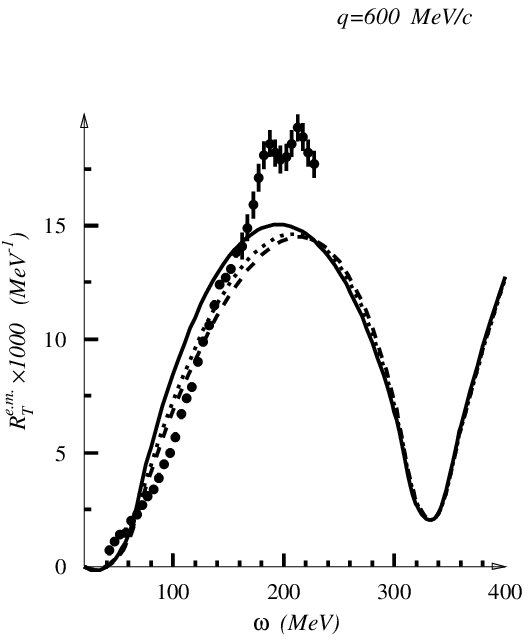,height=7cm}
\end{tabular}
}
\end{center}
\caption{\protect\label{fig9} Mean field calculation for 
the transverse response on $^{12}C$ at different transferred momenta.
Data from \protect\cite{Me-al-84,Me-al-85} (circles) and from 
\protect\cite{Jo-95,Jo-96} (triangles). 
Solid line: Mean field calculation with $m_\rho=600 ~ MeV$;
dashed line: mean field calculation with $m_\rho=770 ~ MeV$;
dotted line: FFG calculation.
}
\end{figure}

\begin{figure}
\begin{center}
\mbox{
\begin{tabular}{cc}
\epsfig{file=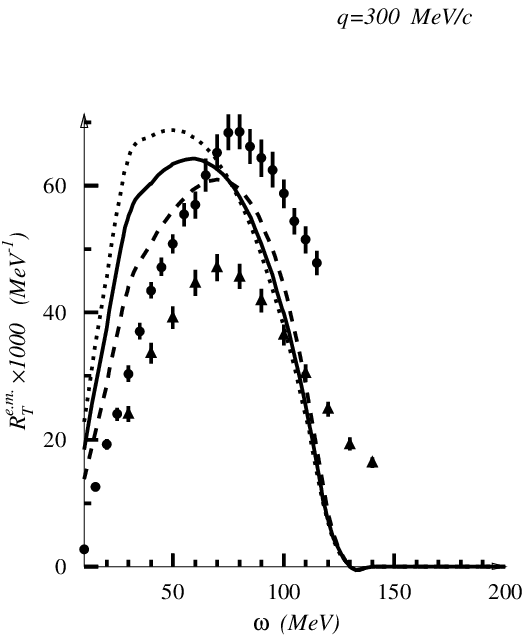}
&
\epsfig{file=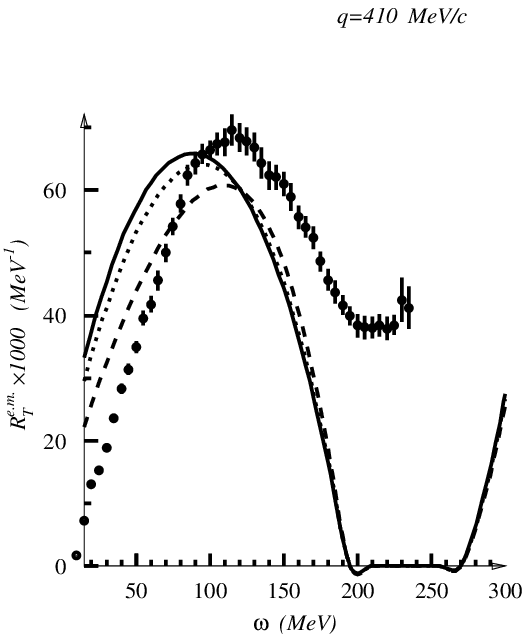}
\cr
\epsfig{file=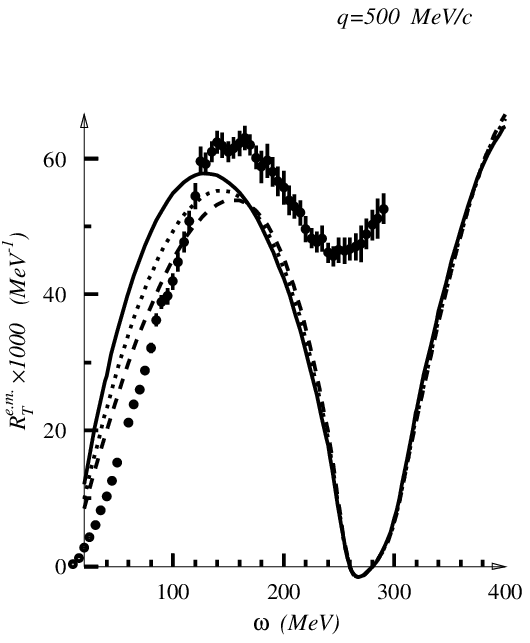}
&
\epsfig{file=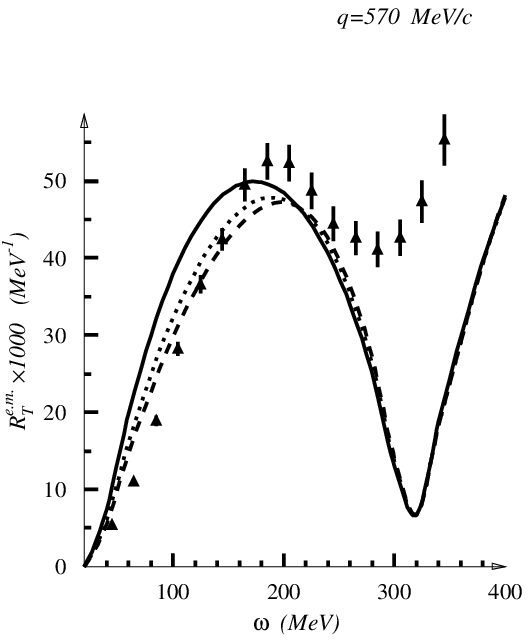}
\end{tabular}
}
\end{center}
\caption{\protect\label{fig10} Mean field calculation for 
the transverse response on $^{40}Ca$ at different transferred momenta.
Data from \protect\cite{Me-al-84,Me-al-85} (circles) and from 
\protect\cite{Jo-95,Jo-96} (triangles). 
Solid line: Mean field calculation with $m_\rho=600 ~ MeV$;
dashed line: mean field calculation with $m_\rho=770 ~ MeV$;
dotted line: FFG calculation.
}
\end{figure}

The effective interaction in the $\rho$ channel is 
shown in fig. \ref{fig8}, together with the one we employed in ref.
\cite{CeCoSa-96} ($m_\rho=770 ~ MeV$, $g^\prime_0=0.5$, 
$q_{c,T}=1.4~GeV/c$, $C_\rho=2.2$, $\Lambda = 2.5~GeV/c$). 
We remark, as already outlined in 
\cite{CeCoSa-96}, that this effective interaction is qualitatively 
analogous to the one used by Oset and coworkers \cite{OsSa-87,CaOs-92}.
The main difference, which is now linked to a phenomenological 
outcome, is the point where the effective interaction changes sign, 
which is, in view of the quoted experimental data, set around 300 
$MeV/c$. It is understood that a refinement of experimental data, 
strongly asked for, hopefully at CEBAF, could lead to a corresponding 
change in our parametrization of the effective interaction.
Still a drawback seems to arise from the used parameters, namely the low 
value of $g^\prime_0$. This is not particularly worrying in the 
transverse channel, but seems to have dramatic consequences in the 
longitudinal one. In fact both $g^\prime_L(q)$ and $g^\prime_T(q)$ must 
have the same Landau limit $g^\prime_0$, and a low value of it should 
immediately lead to the occurrence of a pion condensate.
This is not true, however, because pion condensation, if any, occurs 
typically at momenta of a few hundreds $MeV/c$, where our 
$g^\prime_L(q)$ is sensibly increased with respect to $g^\prime_0$.

Now we come to the comparison between the mean field calculation
and the available data. 
They are displayed in fig. \ref{fig9} for $^{12}C$ and in fig. 
\ref{fig10} for $^{40}Ca$.


Assuming as the most reliable data the one provided by \cite{Jo-95,Jo-96} 
the agreement between mean field and experiment is not completely 
satisfactory, but nevertheless less dramatic than in the case of Saclay 
data. It is clear how the assumption of an effective mass for the $\rho$ 
improves the agreement.

As an aside, without showing the figures for sake of brevity, we observe 
that Iron display the same trend as the Calcium.

\section{The one-loop corrections \label{sect6}}

We finally come to the one-loop corrections, namely the ultimate goal of 
this paper.

First, let us define once forever the sets of parameters we shall use.
We remark however that this same set of parameters is used in all the
calculations presented in this paper.
\begin{enumerate}
\item The mass of the $\rho$-meson has been set to 600 $MeV$ in the 
spin-transverse channel, while in the scalar-isovector has been left 
unchanged ($m_\rho=770~MeV$)
\item The mass of the $\omega$ has been left
unchanged too ($m_\omega=770~MeV$)
\item The pion coupling constant have been assumed, as usual, as
$\frac{f^2_{\pi NN}}{4\pi}=0.08$, $\frac{f^2_{\pi N\Delta}}{4\pi}=0.32$ 
and $\frac{f^2_{\pi \Delta\Delta}}{4\pi}=0.016$. The last value comes 
from the quark model and is of course quite uncertain.
\item We assumed $C_{\rho NN}=C_{\rho N\Delta}=C_{\rho\Delta\Delta}=2.3$
in the spin-transverse channel and $=0.05$ in the scalar-isovector one.
Further, $C_{\omega NN}=C_{\omega \Delta\Delta}=0.15$ in the 
scalar-isoscalar channel and $=1.5$ in the isoscalar spin-transverse one.
\item $g_0^\prime$ is set to 0.35.
\item The many-body cut-off of SRC are put to $q_{c,L}=800 Mev/c$ and
$q_{c,T}=1300 Mev/c$.
\item The pion cut-off at the vertices are set to
$\Lambda_{\pi NN}=1300 Mev/c$, $\Lambda_{\pi 
N\Delta}=\Lambda_{\pi\Delta\Delta}=1000 Mev/c$.
\item The $\rho NN$ cut-off at the vertex in the 
spin-transverse channel is $\Lambda_{\rho NN}=1750 
Mev/c$.
\item All the remaining cut-offs at the vertices are set to $1000 
Mev/c$.
\end{enumerate}

The one-loop corrections to the response stem from many diagrams, 
corresponding to different physical effects.
We start with the longitudinal response and we examine, as a guideline, 
the case of Carbon at a transferred momentum of $300~ MeV/c$.

First let us examine the case of pure nucleon dynamics (no $\Delta$'s) 
and let us also drop the RPA dressing in the mean field (pure Lindhard 
function) as well as in the external legs of the one-loop corrections
(this means that we examine only the contribution stemming from the black 
bubble of fig. \ref{fig2bis}, corresponding to the five diagrams of 
fig. \ref{fig2}). There 
are three contributions there, that we called self-energy, exchange and 
correlation terms. They correspond respectively to the first two diagrams
of the first line of fig.~\ref{fig2}, to the third diagram of the first
line of fig.~\ref{fig2} and to the two diagrams of the second line
of fig.~\ref{fig2}.
With the present choice of the parameters the main 
effects come from the isovector spin-longitudinal and spin-transverse 
channels, that turn out to be rather similar, while the other channels 
give a negligible contribution. In fig. \ref{fig12} these one-loop 
corrections are shown, and clearly, while going in the right direction, 
are still not sufficient to explain the quenching of the peak. In 
passing we observe that exchange and correlation contribution display a trend 
opposite to the one of self-energy diagrams, thus corroborating our 
initial statement that in the loop expansion cancellations are realized
order-by-order.

\begin{figure}
\epsfig{file=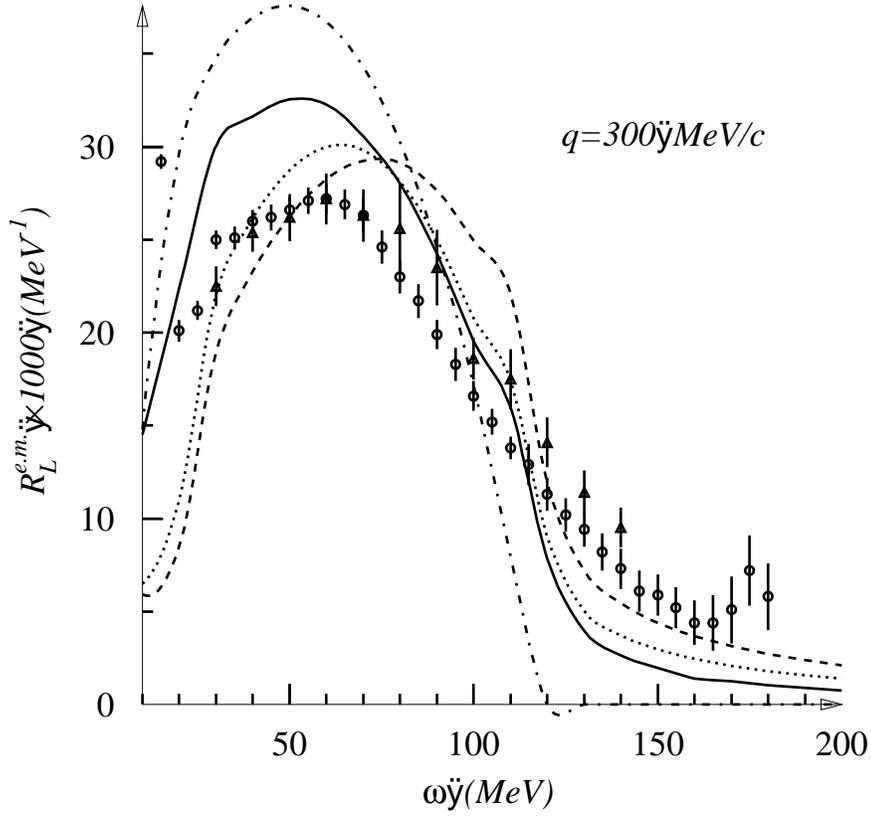}
\caption{\protect\label{fig11}The one-loop corrections to the 
longitudinal response for $^{12}C$ without RPA dressing of the mean 
field and without $\Delta$'s. Dash-dotted line: FFG; dashed line: FFG plus 
self-energy diagrams; dotted line: FFG plus self-energy and exchange 
diagrams; solid line: FFG plus self-energy, exchange and correlation 
diagrams.}
\end{figure}

The next step is to introduce the $\Delta$-resonance. On physical 
grounds we expect that box diagrams will dominate the response. They are 
equivalent to the combination of the subdiagrams 4c) with 4a$^\prime$),
4a with 4c$^\prime$) and 4c with 4c$^\prime$) in fig. \ref{fig5}.
They are shown in fig. \ref{fig12} together with a complete calculation, 
i.e., with all possible diagrams. It appears that the box diagrams are 
dominant indeed and that they make the job the $\sigma$-meson is 
expected to do, namely to strongly enhance the peak. 
There is an important consideration to outline here, namely the fact 
that the peak is now softened.

\begin{figure}
\epsfig{file=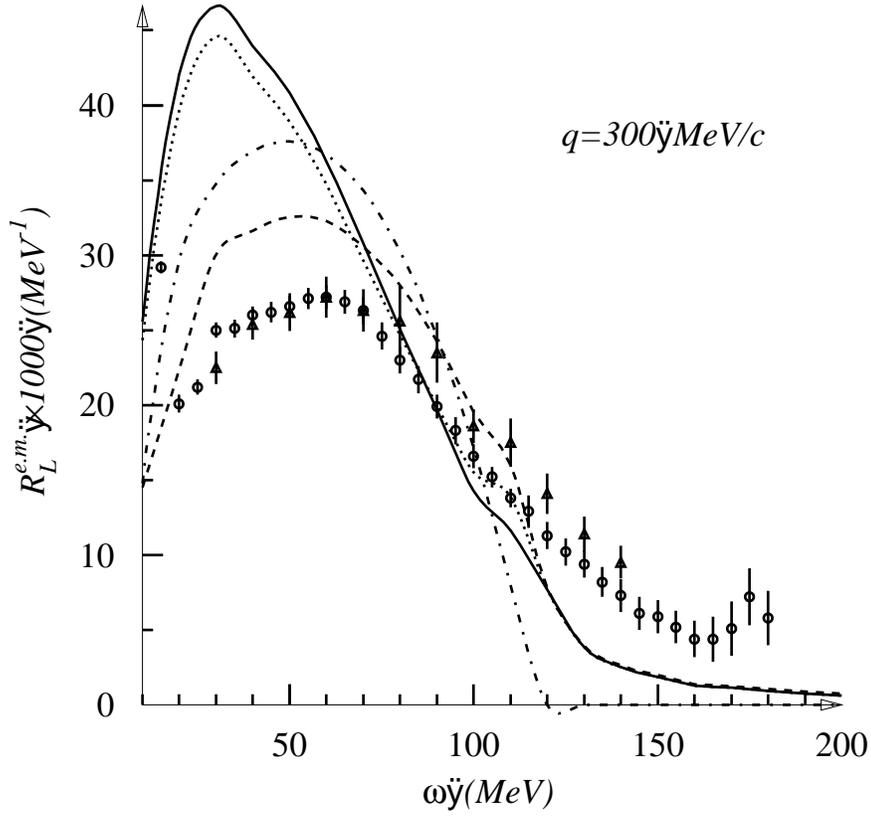}
\caption{\protect\label{fig12}The one-loop corrections to the 
longitudinal response for $^{12}C$ without RPA dressing of the mean 
field. Dash-dotted line: FFG; dashed line: diagrams without $\Delta$;
dotted line: box diagrams only added; solid line: all diagrams.}
\end{figure}

\begin{figure}
\epsfig{file=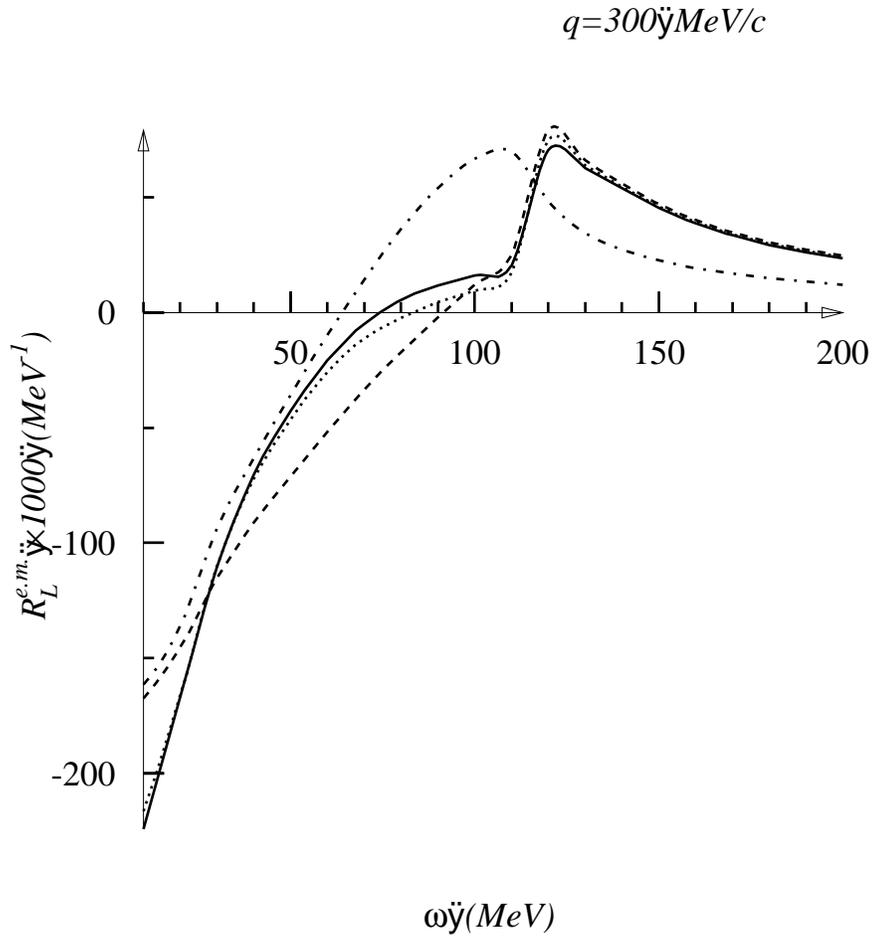}
\caption{\protect\label{fig13}The one-loop corrections to the
real part of the polarization propagator. Diagrams and lines as in fig. 
\protect\ref{fig12}.}
\end{figure}

Since in going to the complete treatment of the response we need the 
real part too of the diagrams, we show in fig. \ref{fig13} the real part 
of the polarization propagator corresponding to the diagrams of fig. 
\ref{fig12}.

\begin{figure}
\epsfig{file=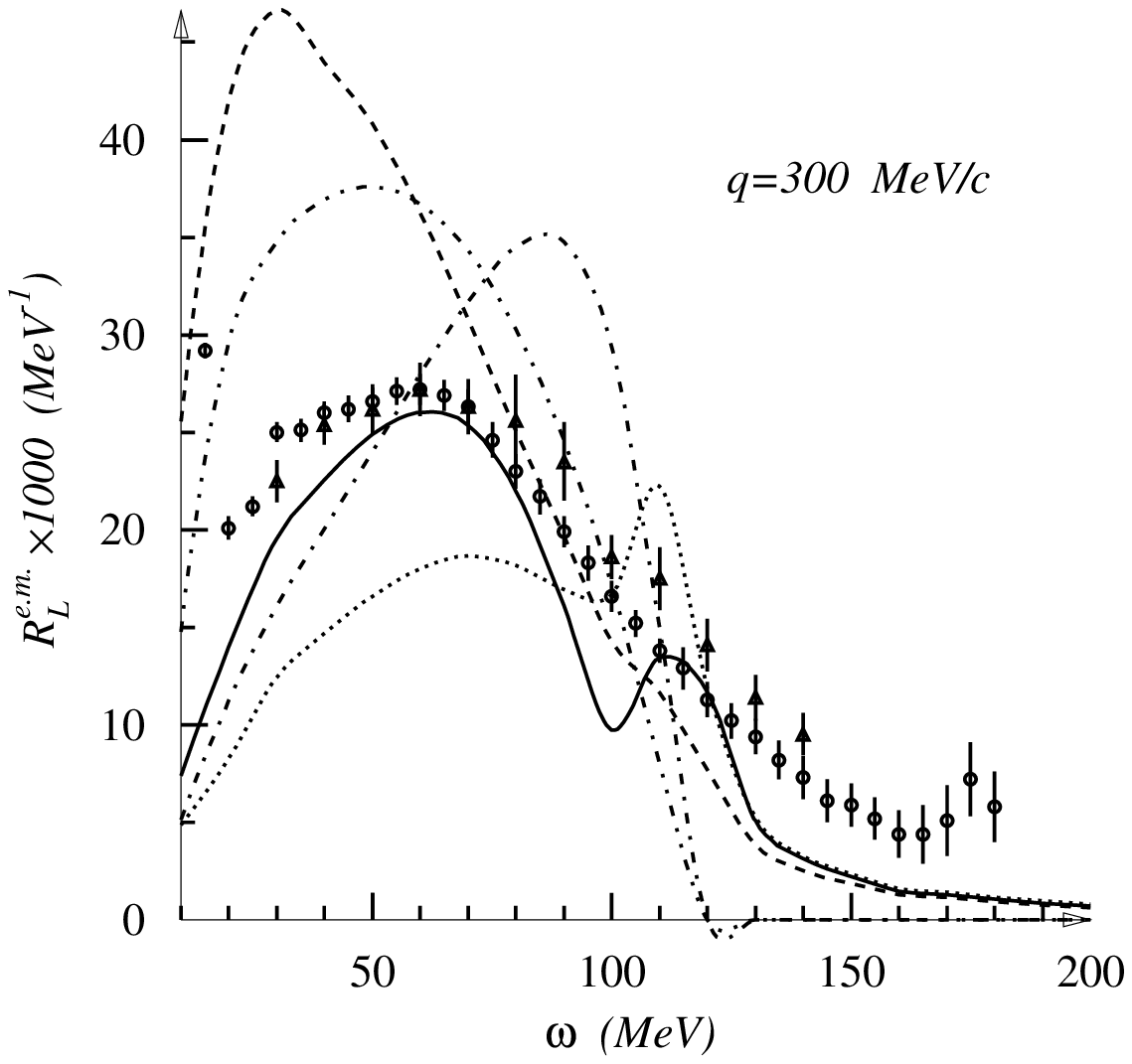}
\caption{\protect\label{fig14} The longitudinal response on $^{12}C$.
Solid line: full response; dashed line: all diagrams without RPA;
dotted line: diagrams without $\Delta$'s only but with RPA; dash-dotted 
line: mean field; dash-dotted-dotted line: 
FFG}
\end{figure}

Finally in fig. \ref{fig14} we plot the complete graph, with the inclusion 
of the RPA dressing everywhere. 
From the figure it is clearly evident the compensation between two different
effects - the hardening coming from the box diagram contributions and the 
softening deriving by RPA dressing at the mean field level: they correspond
in a mean field description to the compensation between $\sigma$- and 
$\omega$-meson, as expected from Dirac phenomenology 
\cite{Wa-74,ClHaMe-82,Cl-86}. The final effect
is to go back to the experimental data.

Next we examine the transverse response, following the same path as 
before. 

\begin{figure}
\epsfig{file=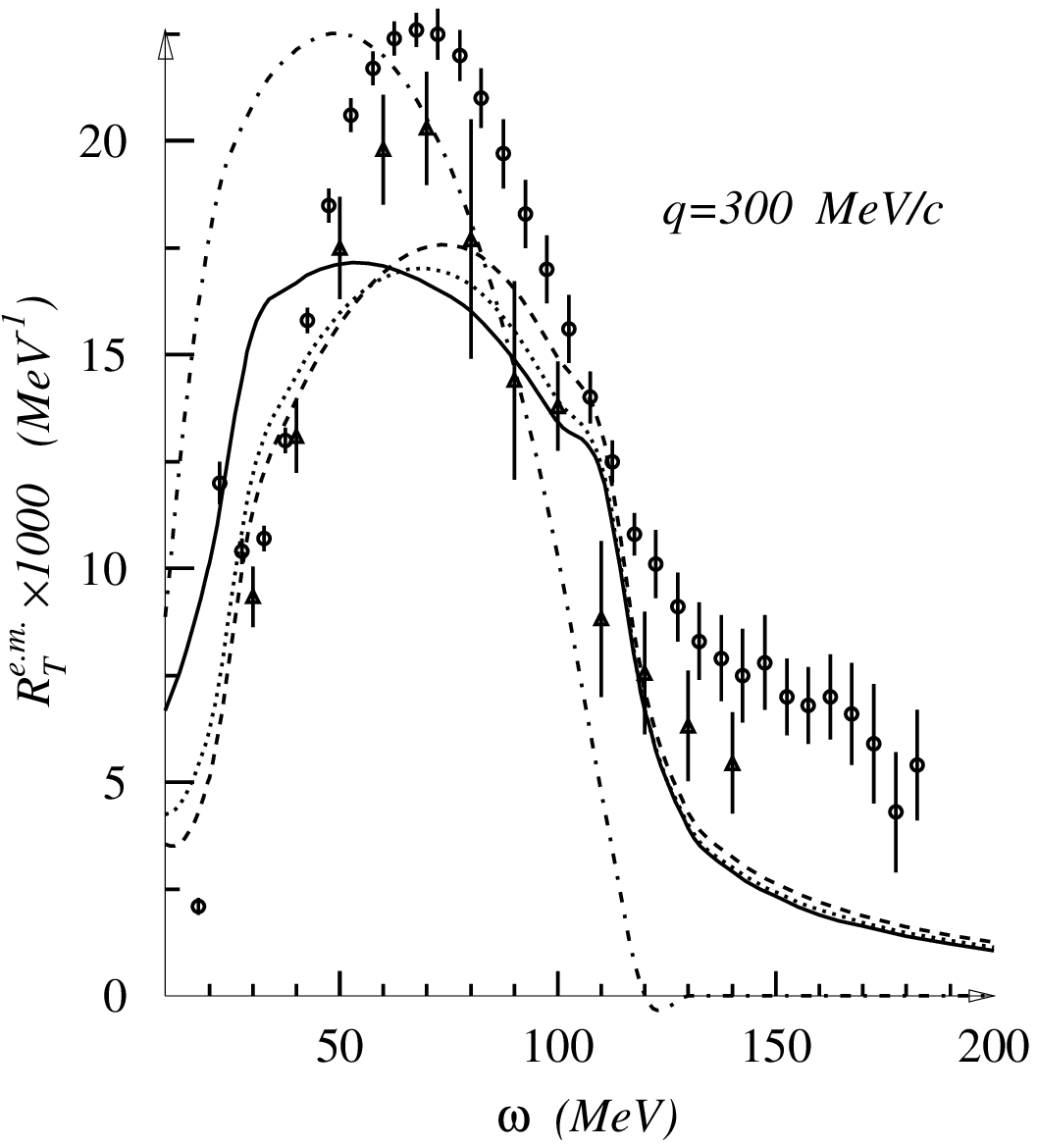}
\caption{\protect\label{fig15}The one-loop corrections to the 
spin-transverse response for $^{12}C$ without RPA dressing of the mean 
field and without $\Delta$'s. Dash-dotted line: FFG; dashed line: FFG plus 
self-energy diagrams; dotted line: FFG plus self-energy and exchange 
diagrams; solid line: FFG plus self-energy, exchange and correlation 
diagrams.}
\end{figure}

In fig. \ref{fig15} the one-loop corrections are drawn, without 
RPA dressing of the mean field and without $\Delta$'s. The path is 
clearly similar to that of fig. \ref{fig11} and is even more 
pronounced, but in the wrong direction, however. This can be understood in 
this way: the main effect, as both figures display, come from the self-energy 
diagrams, that deplete the peak, the only difference being a factor 2 
which emphasizes the depletion in the transverse channel. Thus there is 
no way of reproducing the experimental data (that on the other hand are 
nicely described, up to a shift, 
by the simple FFG curve) without invoking other 
effects, as the $\Delta$ dynamics and the RPA-dressing effects.

\begin{figure}
\epsfig{file=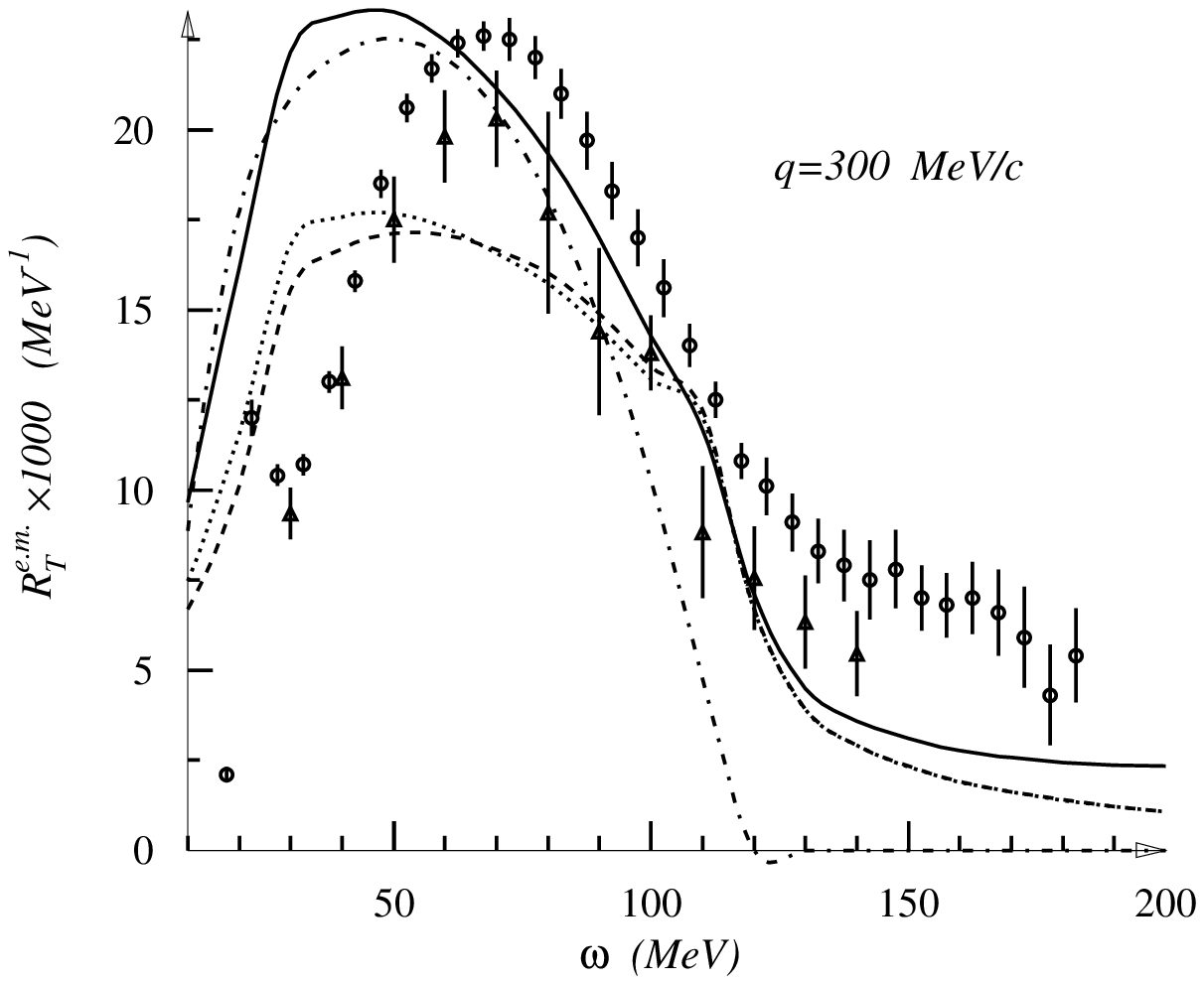}
\caption{\protect\label{fig16}The one-loop corrections to the 
spin-transverse response for $^{12}C$ without RPA dressing of the mean 
field. Dash-dotted line: FFG; dashed line: diagrams without $\Delta$;
dotted line: box diagrams only added; solid line: all diagrams.}
\end{figure}
The contribution of the $\Delta$ diagrams to the transverse response is 
then shown in fig. \ref{fig16}, again without RPA-dressing. This time an 
important difference arises, concerning the contribution of the box 
diagrams, that are almost irrelevant. The origin of this seemingly 
contradictory behaviour stems from the spin traces, shown in table 
\ref{tab7}. It appears there that the spin traces for the case of a 
scalar probe and two spin-transverse particle exchanged are proportional 
to $1+\cos^2\theta_{pq}$ (see appendix \ref{appB} for notations)
while for a spin-transverse probe the traces are proportional to
$\pm \left(\frac{3}{2}-\frac{1}{2}\cos^2\theta_{kq}-\frac{1}{2}\cos^2
\theta_{kp}-\frac{1}{2}\cos^2\theta_{pq}\right)$, thus ensuring large 
cancellation in the latter case, while in the former the two terms sum 
up together. In fig. \ref{fig16} it is evident indeed that the box 
diagrams contribute very poorly to the response, but this time isospin 
selection rule does not forbid any of the remaining diagrams, which 
instead cancel almost completely the depletion induced by the 
self-energy diagrams. In conclusion the one-loop corrections are almost 
negligible in the transverse channel.

\begin{figure}
\epsfig{file=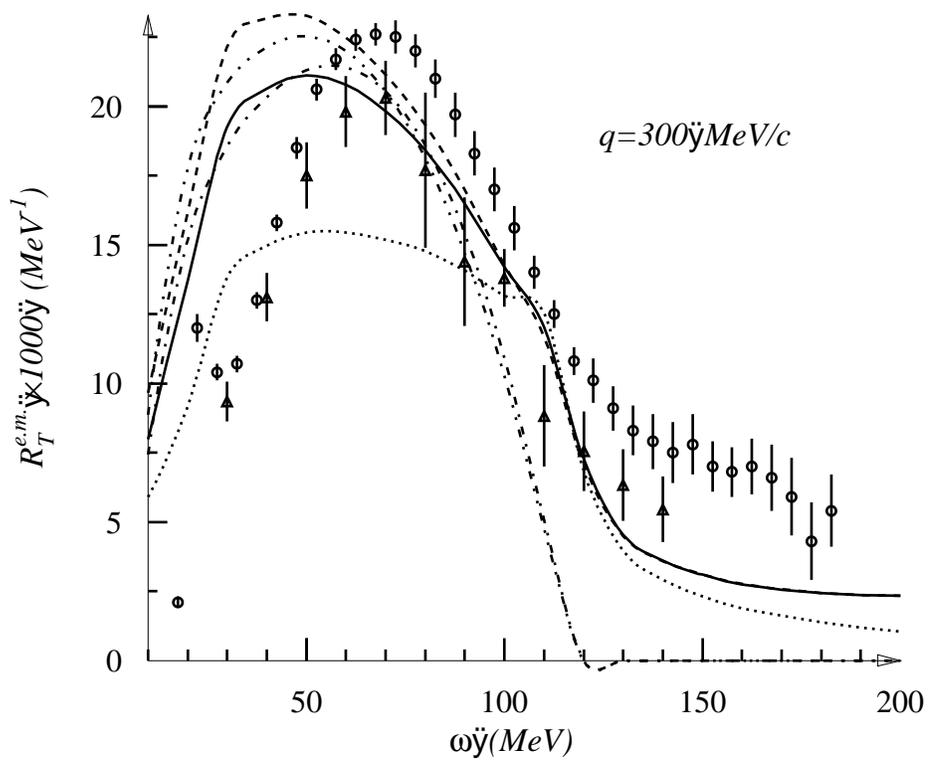}
\caption{\protect\label{fig17} The transverse response on $^{12}C$.
Solid line: full response; dashed line: all diagrams without RPA;
dotted line: diagrams without $\Delta$'s only but with RPA; dash-dotted 
line: mean field; dash-dotted-dotted line: 
FFG}
\end{figure}

Finally in fig. \ref{fig17} the whole calculation is reported.

Here the physical insight are clear: since one-loop corrections are 
negligible, we are now sensitive to the RPA-dressing of the $\rho$-meson 
channel (the dominant one): we require here an effective interaction 
near to vanish, in order to not further deplete the peak. The one we 
have chosen is still a little bit repulsive: clearly the momentum region 
between 300 and 400 MeV/c is the most sensitive to the details of the 
effective interaction. As a matter of fact our results stay more or less 
between Saclay's and Jourdan's data.

\begin{figure}
\begin{center}
\mbox{
\begin{tabular}{cc}
\epsfig{file=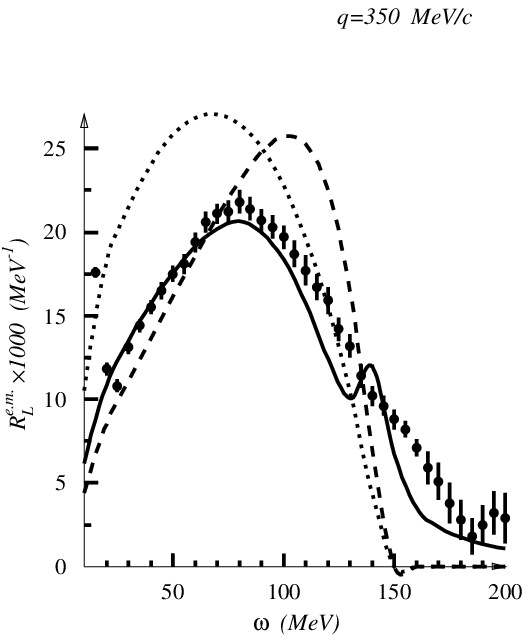,height=5.2cm,width=6cm}
&
\epsfig{file=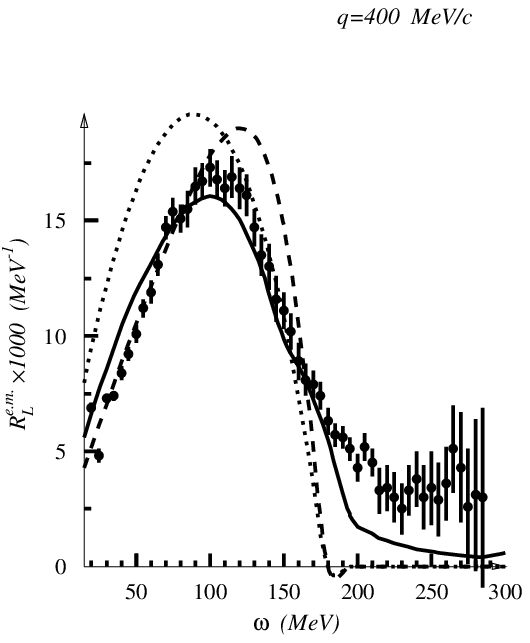,height=5.2cm,width=6cm}
\cr
\epsfig{file=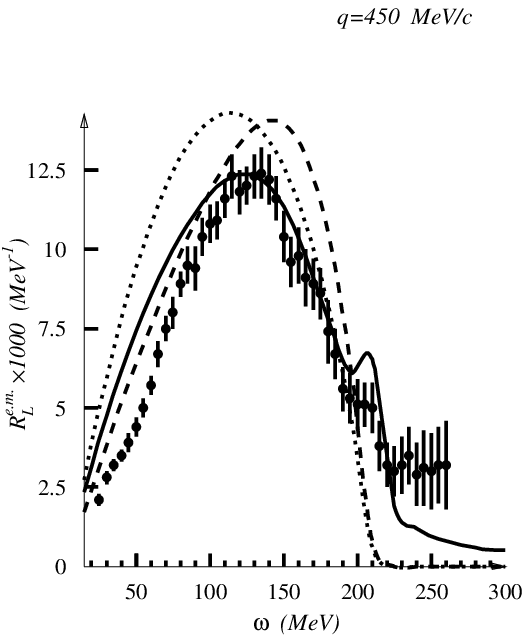,height=5.2cm,width=6cm}
&
\epsfig{file=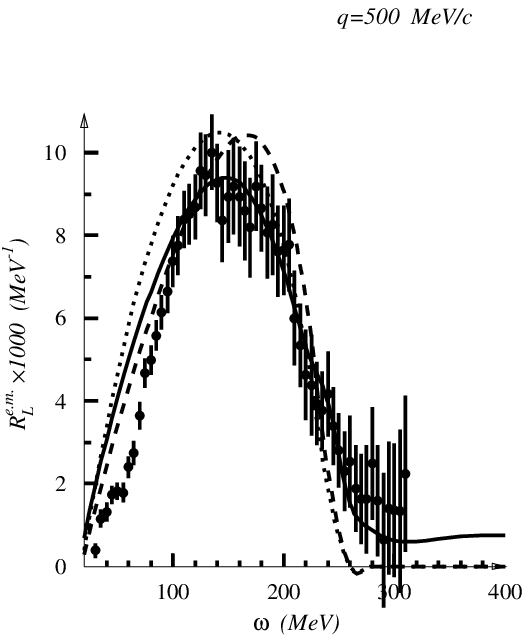,height=5.2cm,width=6cm}
\cr
\epsfig{file=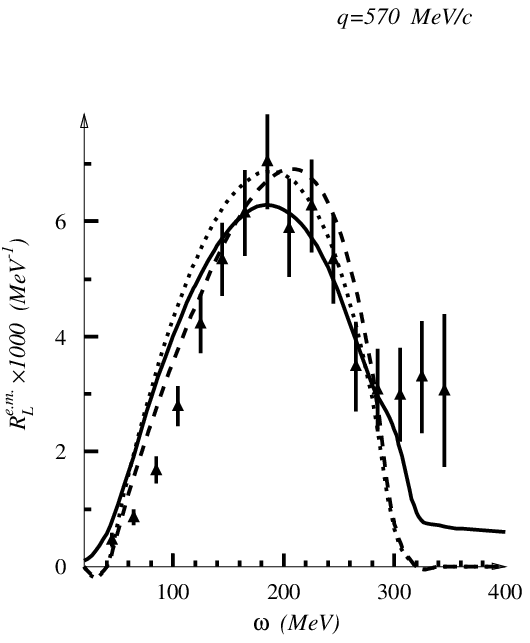,height=5.2cm,width=6cm}
&
\epsfig{file=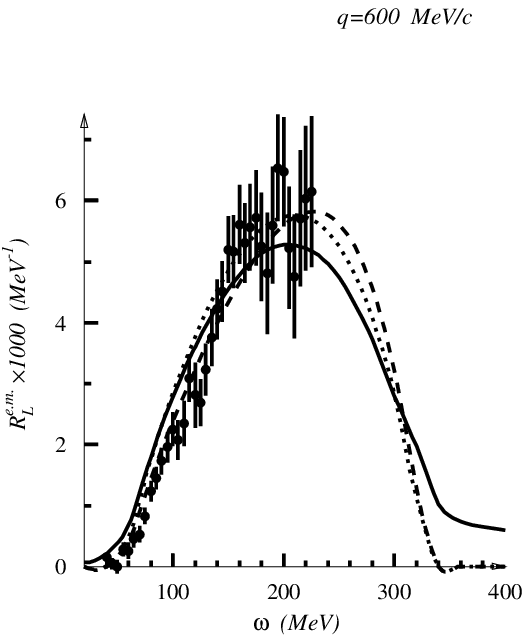,height=5.2cm,width=6cm}
\end{tabular}
}
\end{center}
\caption{\protect\label{fig18} Full calculation for 
the longitudinal response on $^{12}C$ at different transferred momenta.
Data from \protect\cite{Me-al-84,Me-al-85} (circles) and from 
\protect\cite{Jo-95,Jo-96} (triangles). Solid line: full calculation;
dashed line: Mean field 
calculation;
dotted line: FFG calculation.
}
\end{figure}

\begin{figure}
\begin{center}
\mbox{
\begin{tabular}{cc}
\epsfig{file=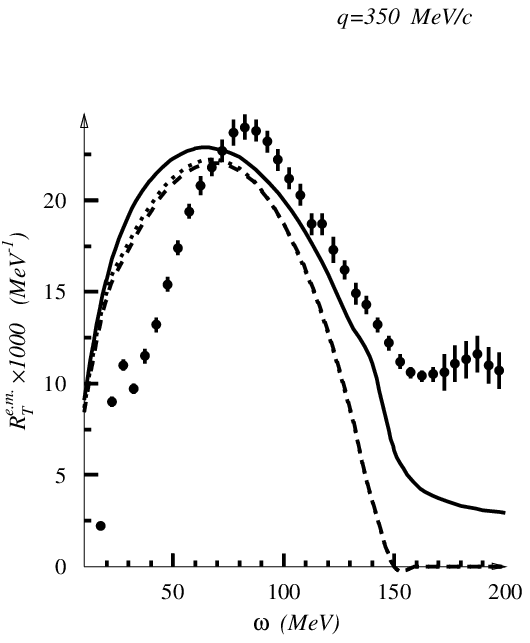,width=6cm}
&
\epsfig{file=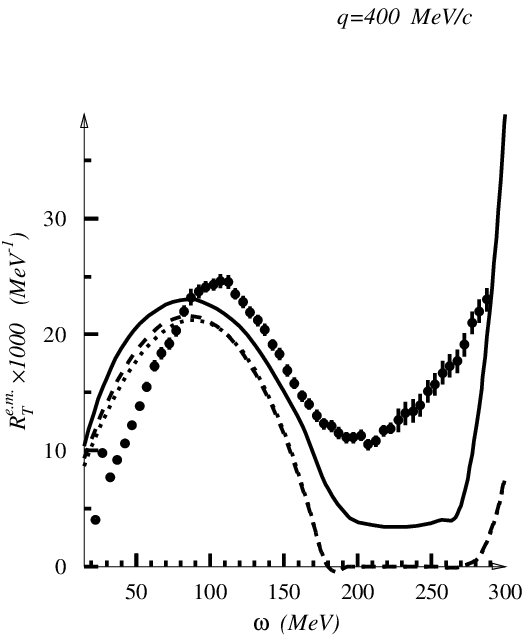,width=6cm}
\cr
\epsfig{file=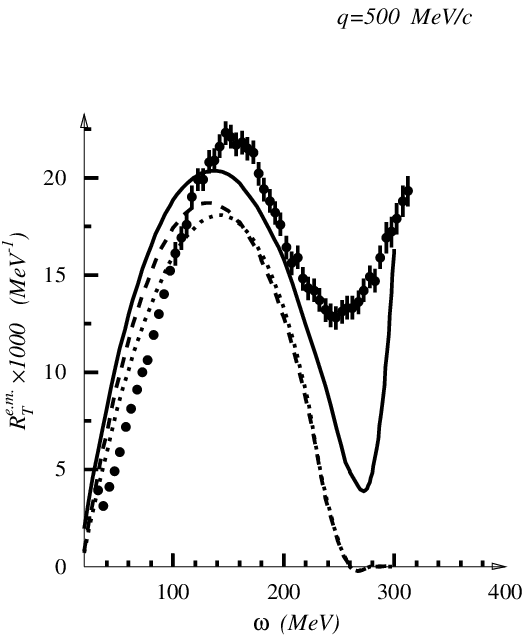,width=6cm}
&
\epsfig{file=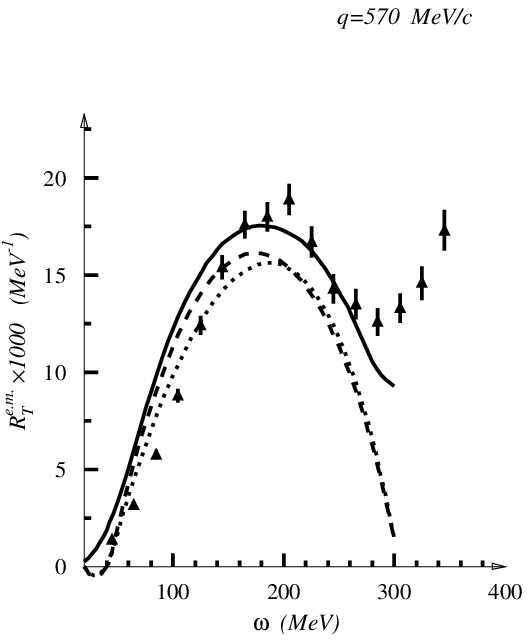,width=6cm}
\end{tabular}
}
\end{center}
\caption{\protect\label{fig19} Full calculation for 
the transverse response on $^{12}C$ at different transferred momenta.
Data from \protect\cite{Me-al-84,Me-al-85} (circles) and from 
\protect\cite{Jo-95,Jo-96} (triangles). Solid line: full calculation;
dashed line: Mean field 
calculation;
dotted line: FFG calculation.
}
\end{figure}

A more complete survey of our results for $^{12}C$ is presented in fig. 
\ref{fig18} for the longitudinal and in fig. \ref{fig19} for the 
transverse case.

\begin{figure}
\begin{center}
\mbox{
\begin{tabular}{cc}
\epsfig{file=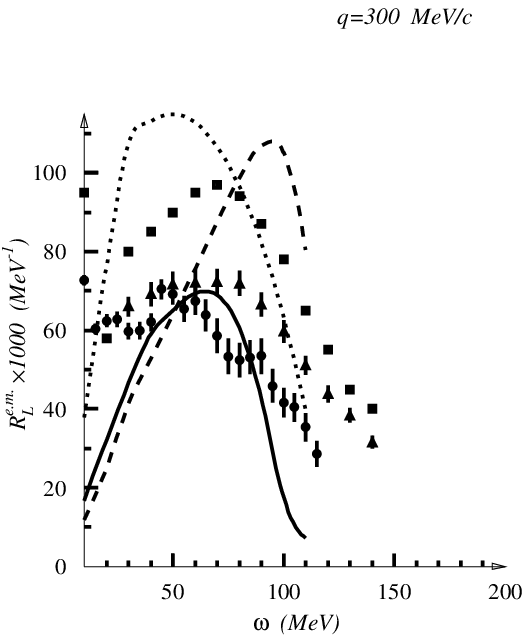,width=6cm}
&
\epsfig{file=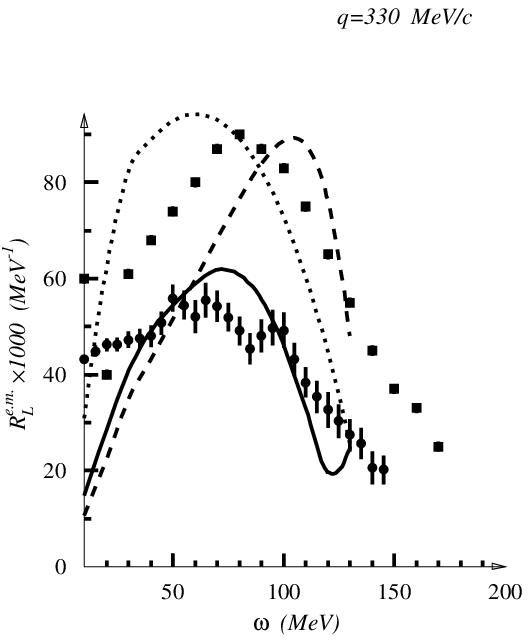,width=6cm}
\cr
\epsfig{file=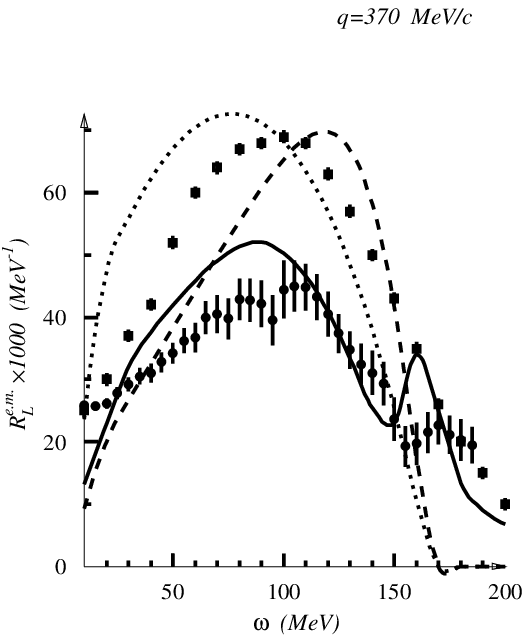,width=6cm}
&
\epsfig{file=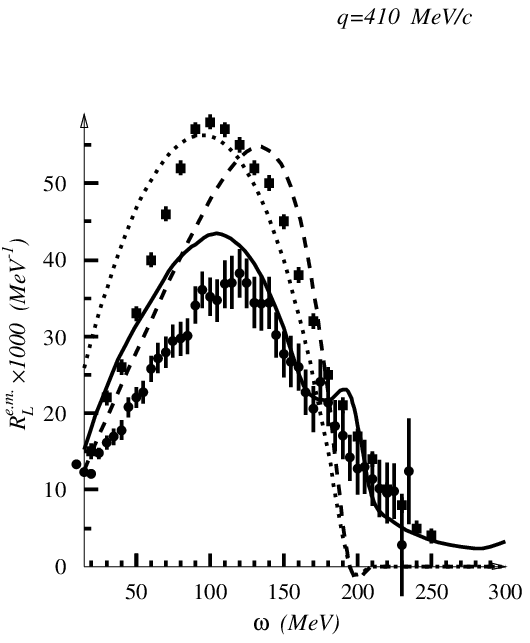,width=6cm}
\end{tabular}
}
\end{center}
\caption{\protect\label{fig20} Full calculation for 
the longitudinal response on $^{40}Ca$ at different transferred momenta.
Data from \protect\cite{Me-al-84,Me-al-85} (circles) and from 
\protect\cite{Jo-95,Jo-96} (triangles). Solid line: full calculation;
dashed line: Mean field 
calculation;
dotted line: FFG calculation.
}
\end{figure}

\begin{figure}
\begin{center}
\mbox{
\begin{tabular}{cc}
\epsfig{file=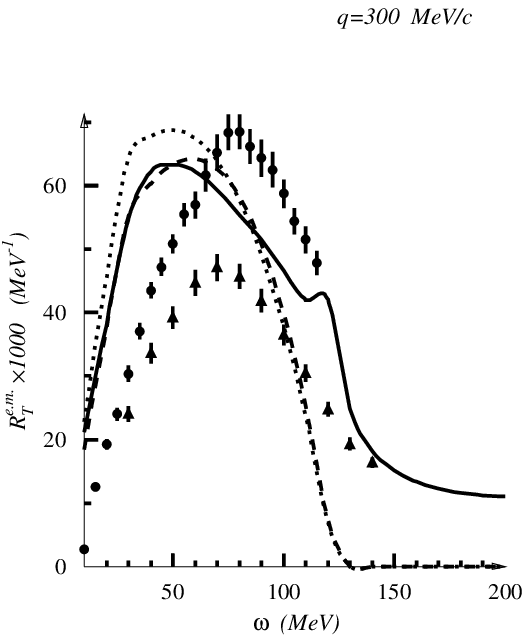,height=5.2cm,width=6cm}
&
\epsfig{file=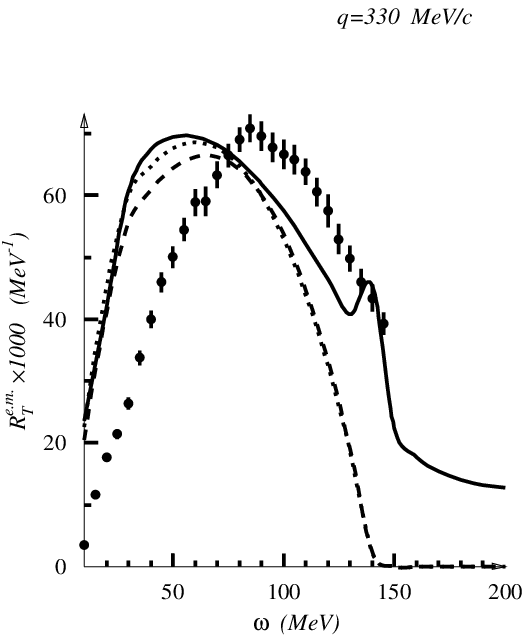,height=5.2cm,width=6cm}
\cr
\epsfig{file=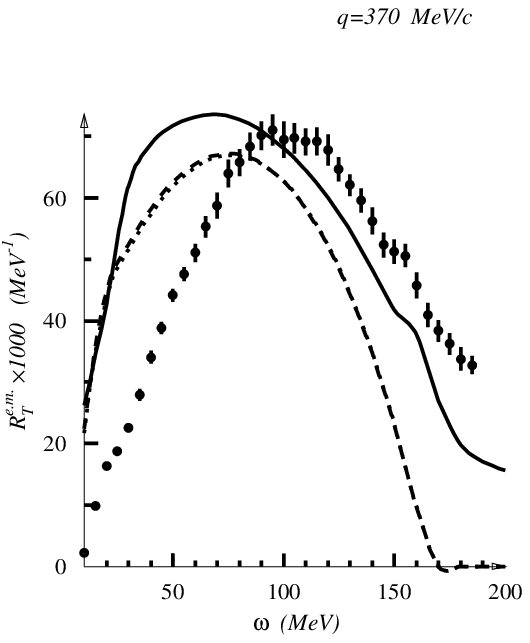,height=5.2cm,width=6cm}
&
\epsfig{file=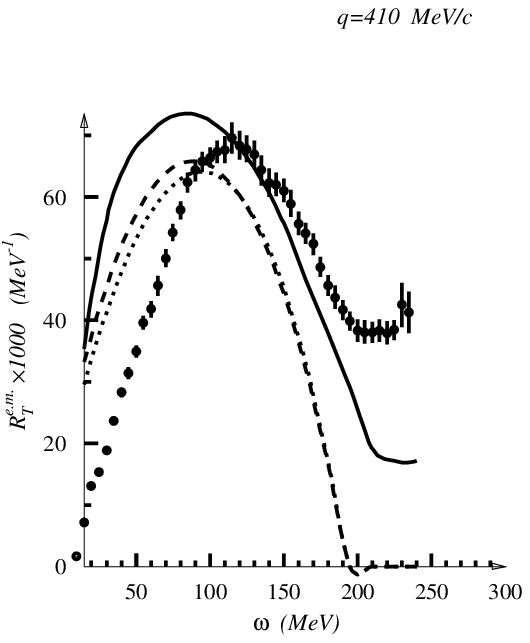,height=5.2cm,width=6cm}
\cr
\epsfig{file=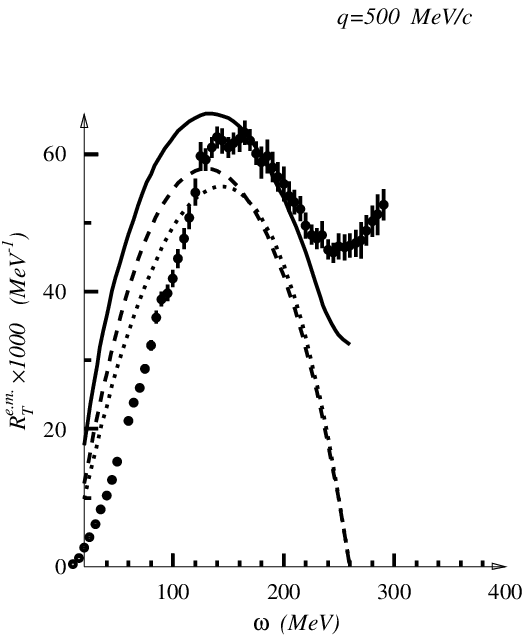,height=5.2cm,width=6cm}
&
\epsfig{file=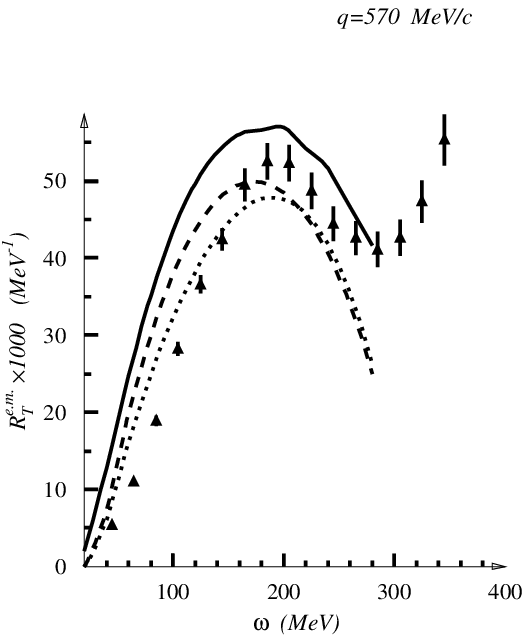,height=5.2cm,width=6cm}
\end{tabular}
}
\end{center}
\caption{\protect\label{fig21} Full calculation for 
the transverse response on $^{40}Ca$ at different transferred momenta.
Data from \protect\cite{Me-al-84,Me-al-85} (circles) and from 
\protect\cite{Jo-95,Jo-96} (triangles). Solid line: full calculation;
dashed line: Mean field 
calculation;
dotted line: FFG calculation.
}
\end{figure}

The figures \ref{fig20} and \ref{fig21} present instead the results of 
our calculation ons $^{40}Ca$. In this case we have assumed an higher 
value for $\k$, namely $\k$=1.2~fm$^{-1}$, which better describes a 
medium nucleus. The interplay between RPA dressing of the external legs 
of the diagrams and the one-boson-loop corrections, which also alter the 
real part of the diagram (a density-dependent effect) leads to a more 
pronounced depletion of the longitudinal response, in agreement with 
the Saclay data. This results is relevant in our opinion and deserves a 
comment. First of all we stress once more that our approach is
semi-phenomenological and that we have adjusted the free parameters of 
the model in such a way to agree with the Saclay data and other more 
refined experimental results are expected before to draw definite 
conclusions. For a long time the apparent discrepancy between the Saclay 
data on Carbon and Calcium, that provide so different Coulomb sum rule, 
led some people to question about the Calcium data.
We have shown here that such a discrepancy {\em may} be explained as a density 
effect in the frame of a well defined theoretical model (the boson loop 
expansion), without violating the Coulomb sum rule. In fact it is clear 
from figs. \ref{fig17} and \ref{fig19} that the one-loop corrections are 
washed out when the transferred momentum increases and since the 
effective interaction in the spin-scalar channels is also vanishing at 
high $q$ then the mean field result also reduces in this limit
to the FFG result: then the Coulomb sum rule is automatically fulfilled .

Still our results display a drawback. In the $^{40}Ca$ data ad low 
transferred momenta we have not displayed our results in the tail of the 
QEP. There an instability is already apparent in the $^{12}C$ data 
(where it could be regularized by interpolation procedure used in 
drawing the curves) and 
becomes more pronounced in Calcium. This could be unpleasant, but it is in 
our opinion consistent with our approach. The main difficulty of the 
boson loop expansion is that it is impossible at a given finite order in 
the number of loops to account for the shift of the QEP. This comes out 
in fact from the sum of the whole Dyson series in the fermionic lines of 
the particle-hole propagator, that in turn amounts to consider an 
infinite order in the loop expansion. The self-energy term in the 
one-boson loop correction tries to simulate the shift of the QEP by 
reshaping it, so that this diagram is particularly effective on the 
edges of the peak. On the other hand we know that our approach preserves 
all general theorems and sum rules (like, e.g., the $f^\prime$ sum rule) 
thus forcing, on the whole, a cancellation between different 
contributions, and in fact exchange and correlation terms have opposite 
sign with respect to the self-energy. The cancellation is ensured 
however at the level of the integrated response, while in the details 
some mismatches can survive were the corrections are more pronounced. 
Coherently with our approach this happens exactly where the 
one-boson-loop approximation must fail, i.e., on the edges of the QEP 
where its shift cannot be accounted for. Remarkably the RPA dressing of 
the external legs emphasizes the effect.

\begin{figure}
\centerline{
\epsfig{file=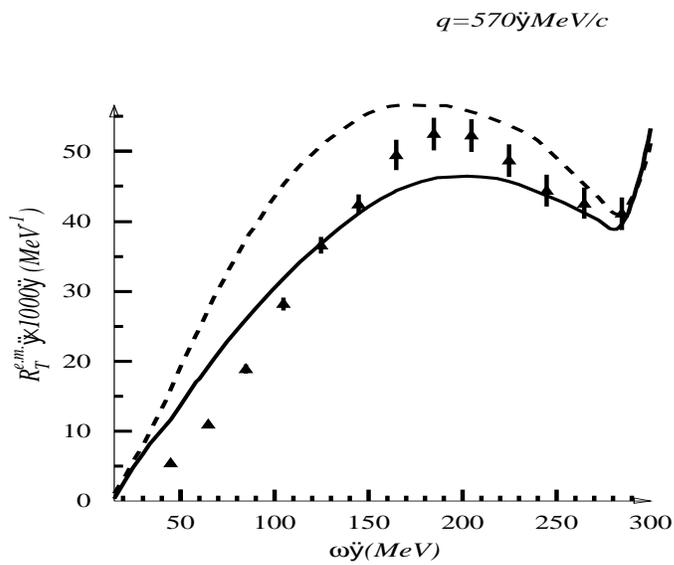,height=8cm,width=10cm}
}
\caption{\protect\label{fig22}Transverse response for $^{40}Ca$ with all 
diagrams with different $\rho$-masses. Solid line: $m_\rho=770$ MeV,
dashed line: $m_\rho=600$ MeV.}
\end{figure}

Finally, we want to discuss the discrepancies between Saclay and Jourdan 
data. These are particularly emphasized in the transverse response. 
Recalling our previous discussion, the Jourdan data seem to require a 
less pronounced attraction in the isovector spin-transverse effective 
interaction. Since we decreased in our calculations the $\rho$-mass just
to emphasize the attractive part of the interaction we reproduce in fig. 
\ref{fig22} our results both with $m_\rho$=770 MeV and with $m_\rho$=600
MeV. The figure seems to suggest that a value more or less in between 
the two we have displayed should
better agree with the Jourdan data.
Considering however that still the uncertainty in the experimental 
situation survives and that no microscopical calculations are presently 
available for the $\rho$-meson mass in the nuclear medium, further 
discussions on this topic are still premature.

\section{\label{sect7} Conclusions}

Having presented in the previous section our numerical results we can 
now try to critically examine our approach and to draw some conclusions.

As emphasized in sect. \ref{sect3} we changed somehow our philosophy 
during the completion of this program and ultimately we arrived to see 
our approach as an improvement upon the Landau-Migdal theory of nuclear 
matter.

As is well known, in fact, concerning the nuclear responses, the 
Landau-Migdal theory amounts to parametrize (usually with a constant) 
the effective interaction in a given channel and to explain the 
responses in terms of RPA, nothing but our mean field level.

The boson loop expansion first of all embodies the Landau-Migdal theory 
in a well behaved theoretical frame, thus opening the way to explore its 
microscopical content. 
In the present approach however we limited ourselves still 
to a semi-phenomenological theory by fitting the input in such a way to 
reproduce experimental outcome, but setting it at the one-loop order 
instead of at the mean field level.

The advantages of this procedure are physically relevant. As already 
explained the one-loop corrections are in fact sensitive to the momentum 
dependence of the effective interaction, while the mean field is not (at 
least up to a large extent). Furthermore degrees of freedom like 
$\Delta$-resonances can be explicitly handled in the internal lines, 
their contribution being relevant, as our results show.

Further, all sum rules are preserved at the one-loop order. In 
particular our results show that at sufficiently high $q$ (of the order 
of 600 MeV) no significant discrepancy from Coulomb sum rule is 
observed. $f^\prime$ sum rule is not examined here, but we will devote 
to it a future paper. As noted in ref. \cite{CeSa-94} however, 
with a poorer dynamical saturation of $f^\prime$ sum rule already 
occurred.

The phenomenological input of our model is represented by the effective 
particle-hole interactions. They follow, as far as possible, the 
standard schemes of meson-exchange models, but contain a parameter, 
stemming from many-body mechanisms, that fixes their range and is of 
central importance in ruling the size of the one-loop corrections. We 
have seen that the most relevant effective interaction, namely the 
isovector spin-transverse one, is also almost unknown. However, the 
one-loop corrections in this channels are so weak that the experimental 
data can be directly used to fix it.

The detailed analysis given above shows then a completely different 
dynamics in the longitudinal and transverse channels.
The longitudinal channel is heavily affected by the $\sigma$-$\omega$ 
(the former being described by the box diagrams)
interplay, that not only depletes, but also reshapes the peak, while the 
transverse one by one side is unaffected by $\sigma$ and $\omega$ 
exchange, and on the other displays major corrections coming from the 
direct $\Delta$-excitations, that tends to cancel all the one-loop 
effects.

It is obvious that the full determination of the effective interaction
is a task that can be done only when the approximation scheme is given and
that, in this sense, the present work could be seen as an essentially more
complicated fit to the effective interaction itself. This is partly true,
at least because we must redetermine the parameters at the
1-loop order. But the critical point is that this procedure could not be
carried on at the mean field level, as it was shown in sect.~\ref{sect5},
or, at least, it cannot be performed with reasonable choices for the 
parameters determining the effective interaction: this means, conversely,
that the 1-loop corrections introduced into the calculation some non trivial 
and channel dependent momentum and/or energy dependences into the responses;
these behaviours cannot be convincingly explained (or simulated) at the mean 
field level. We distinguish here two different regimes.
First, we repeatedly stated that the most critical parameters 
are those ruling the medium/high momentum behaviour of the effective
interaction, namely are $q_{c\,L}$
and $q_{c\,T}$, i.e., exactly those parameters which cannot be fixed at the
mean field level. Secondly , the low momentum parts of the effective 
interaction is ruled by the Landau-Migdal parameters, and they are 
conceptually different at the 
mean field level and at 1-loop level. In the former case in fact they 
account also for the non-nucleonic dynamics, while in the latter the box 
diagrams have been disentagled from the Landau-Migdal parameters and 
explicitly accounted for. As a consequence the contrast between the 
value of the Landau-Migdal parameters and the corresponding microscopic 
interactions is dramatic at the mean field level, while is qualitatively 
weakened in going to the 
1-loop level -- an encouraging achievement.

In this context a seemingly disturbing feature is
the value of $g^\prime(0)$, which turns out to be .35; we want 
to outline that a conventional value like $g^\prime(0)=g^\prime_0\simeq .7$ was
primarily motivated by the need of preventing the unwanted and 
unobserved pion condensation: this requirement truly fixes a
lower bound on  $g^\prime(q\simeq 2k_F)$, which coincides with
$g^\prime_0$ only in the pure Landau-Migdal theory. In our momentum
dependent description of $g^\prime_L$ we obtain, in fact,
$g^\prime(q\simeq 2k_F)\simeq 0.7$.

Having so characterized our approach it is now clear why we have adapted 
our parameters on the Saclay's results: simply it was the only complete 
set of data available. When the transverse response from Bates will 
become available too, the same procedure can be repeated and reasonably 
a different effective interaction could be found able to explain these 
data. Clearly this possibility does not alter our theoretical frame. 
Needless to say, the physical information will change, because 
ultimately we can interpret the search for the parameters of our model 
as a way to get information about the effective interactions. Thus we 
complain once more the present ambiguities of the experimental data, 
because it precludes the possibility of attributing a safe
physical meaning to the effective interaction derived in this work.
To further exemplify this topic, let us remind that while Saclay data 
seem to suggest an in-medium-mass for the $\rho$-meson around 600 $MeV$,
the analyses of Jourdan are more compatible with a somehow higher 
$\rho$-mass.
Again this outcome stresses the relevance of a reliable set of 
experimental data.

To conclude this paper, we also want to outline the persisting 
weaknesses of our approach. It is apparent in fact that the transverse 
response comes out to be too large with respect to experimental data
and, if one looks carefully, the same seems to be true for the 
longitudinal one at high transferred momentum. Furthermore the 
$q$-dependence of the corrections could be somehow overemphasized, as 
they seems to be more pronounced at low $q$'s and to fade out too 
quickly (not dramatically, however). This last topic suggests a careful 
analysis of the nuclear densities we have chosen. In fact, for 
dimensional reasons, the fading out of the corrections is expected to 
vary with $k_F/q$ and a lower value of $k_F$ could both narrow the peak 
and spread over a larger range the fading of the 1-loop corrections.

We used in this paper $k_F=1.1 fm^{-1}$ for $^{12}C$
and $k_F=1.2 fm^{-1}$ for $^{40}Ca$, very traditional figures for the
Fermi momentum of medium/light nuclei. 
They come however from old analyses carried out at rather low momenta. 
Many accidents could make this choice questionable: the relativistic 
kinematics first of all, and then the need in light nuclei of a local 
density approach (equivalent in some sense to a lower average density) 
an so on. Last but least the effect of the $\Delta$ tail in the QEP 
region should also be accounted for before extracting from QEP a 
phenomenological value for $k_F$.

All these considerations suggest, before going on with an improved 
dynamics (we mean at least energy-dependent meson propagators), to 
carefully re-examine the responses in the region where, as our calculations 
show, the FFG is quite sufficient to explain the data, i.e, above -- 
roughly -- 600 $MeV/c$. Thus a non-trivial task, to be pursued in a near 
future, should be a careful re-derivation of the FFG parameters in the 
relativistic regime. In these conditions it is known \cite{CeDoMo-97-t} 
that the nuclear and nucleon physics are inextricably linked. Curiously 
enough, the solution of this last problem becomes a fundamental input 
for the  models that should explain the dynamics at lower energy and 
momentum domains.

\appendix

\section{The matrices $\bfm{\cal S}$ \label{appA}}

The matrices $\bfm{\cal S}$ are defined as
\begin{eqnarray}
&{\cal S}_1=\left(\matrix{0&\sqrt{3}&0&0\cr\sqrt{3}&0&2&0\cr
0&2&0&\sqrt{3}\cr0&0&\sqrt{3}&0}\right)\qquad
{\cal S}_2=\left(\matrix{0&-i\sqrt{3}&0&0\cr i\sqrt{3}&0&-2i&0\cr
0&2i0&-i\sqrt{3}\cr0&0&i\sqrt{3}&0}\right)&\nonumber\\
&{\cal S}_3=\left(\matrix{3&0&0&0\cr 0&1&0&0\cr 0&0&-1&0\cr
0&0&0&-3}\right)&
\label{lt11}
\end{eqnarray}

The traces we need are listed below:
\begin{eqnarray}
\Tr(\sigma_i\sigma_j)&=&2\delta_{ij}\\
\Tr(\sigma_i\sigma_j\sigma_k)&=&2i\epsilon_{ijk}\\
\Tr(\sigma_i\sigma_j\sigma_k\sigma_l&=&2(\delta_{ij}\delta_{kl}
-\delta_{ik}\delta{jl}+\delta_{il}\delta_{jk})\\
\Tr(S^\dagger_i S_j)&=&\frac{4}{3}\delta_{ij}\\
\Tr(S^\dagger_i S_j\sigma_k)&=&-\frac{2}{3}i\epsilon_{ijk}\\
\Tr(S^\dagger_i {\cal S}_j S_k)&=&\frac{10}{3}i\epsilon_{ijk}\\
\Tr(\sigma_i\sigma_j S^\dagger_k S_l)&=&\frac{4}{3}\delta_{ij}\delta_{kl}
+\frac{2}{3}\delta_{ik}\delta{jl}-\frac{2}{3}\delta_{il}\delta_{jk}\\
\Tr(S^\dagger_i S_j S^\dagger_k S_l)&=&\frac{8}{9}\delta_{ij}\delta_{kl}
-\frac{2}{9}\delta_{ik}\delta{jl}+\frac{2}{9}\delta_{il}\delta_{jk}\\
\Tr(S^\dagger_i {\cal S}_j S_k\sigma_l)&=&-\frac{2}{3}\delta_{ij}\delta_{kl}
+\frac{8}{3}\delta_{ik}\delta{jl}-\frac{2}{3}\delta_{il}\delta_{jk}\\
\Tr(S^\dagger_i {\cal S}_j {\cal S}_k  S_l)&=&-\frac{2}{3}\delta_{ij}\delta_{kl}
-\frac{22}{3}\delta_{ik}\delta{jl}+\frac{28}{3}\delta_{il}\delta_{jk}
\end{eqnarray}

\section{Evaluation of the traces \label{appB}}

The traces of the diagrams are composed by a spin and an isospin part. 
Further, we must examine the diagrams 1, 3 and 4 (the traces of the 
diagrams 2 and 5 are equal to those of 1 and 4).
To fix the kinematic, we denote with $\bf k$ the momentum of the probe,
and with $\bf q$ the integrated momentum flowing through the bosonic loop
and with $\bf p$ the sum ${\bf p}={\bf k}+{\bf q}$. Further, we denote 
with $\theta_{xy}$ the angles between the vectors $\bf x$ and $\bf y$.
The spin structure of the vertices is assumed to be $1$ for a scalar 
interaction, $\bfm{\hat{h}}\cdot \bfm\sigma$ for a spin-longitudinal 
interaction and $\bfm{\hat{h}}\times\bfm{\sigma}$ 
for a spin-transverse interaction,
$\bf h$ being the momentum entering the vertex.

\subsection{Diagrams 1 and 2}

\subsubsection{Isospin part}

The coefficient are resumed in the table \ref{tab1}. The first column 
labels the diagrams as in fig. \ref{fig3}. SS denotes isoscalar probe 
and isoscalar interaction, SV denotes isoscalar probe
and isovector interaction, VS denotes  isovector probe and isoscalar 
interaction, VV denotes  isovector probe and and isovector interaction.
\begin{table}
\centerline{
\begin{tabular}{||c||c|c|c|c||}
\hline\hline
~&SS&SV&VS&VV\cr
\hline\hline
a&2&6&2&6\cr
\hline
b&0&0&4/3&4\cr
\hline
c&0&4&0&4\cr
\hline
d&0&0&0&8/3\cr
\hline
e&0&0&0&4/3\cr
\hline
f&0&0&0&4/3\cr
\hline
g&0&4&0&20\cr
\hline\hline
\end{tabular}
}
\caption{\protect\label{tab1} Isospin coefficients of the diagrams of fig.
\protect\ref{fig3}}
\end{table}

\subsubsection{Spin part}

For the spin part the coefficients are given in table \ref{tab2}.
Their analytic structure is $\alpha+\beta \cos^2\theta_{kq}$
Diagrams are labelled as in table \ref{tab1}. On the top row the first 
letter characterizes the probe and the second the interaction, with the 
convention: S = scalar channel, L = spin-longitudinal channel, T = 
spin-transverse channel. In each entry the quantity $\alpha$ is reported.
If $\beta$ is non-vanishing, it is also reported after a semicolon.
Finally, the in the spin-transverse channel the trace is taken only over 
the spin part of the diagram and not on its spatial part, which still 
keeps the form $\delta_{ij}-k_i k_j/k^2$. The spatial traces then carry an 
extra factor 2.

\begin{table}
\centerline{
\begin{tabular}{||c||c|c|c|c|c|c|c|c|c||}
\hline\hline
~&SS&SL&ST&LS&LL&LT&TS&TL&TT\cr
\hline\hline
a&2&2&4&2&2&4&2&2&4\cr
\hline
b&0&0&0&4/3&4/3&8/3&4/3&4/3&8/3\cr
\hline
c&0&4/3&8/3&0&4/3&8/3&0&4/3&8/3\cr
\hline
d&0&0&0&0&8/9&16/9&0&8/9&16/9\cr
\hline
e&0&0&0&0&2/9;2/3&10/9;-2/3&0&5/9;-1/3&7/9;1/3\cr
\hline
f&0&0&0&4/3&28/3;-8&32/3;8&4/3&16/3;4&44/3;-4\cr
\hline
g&0&4/3&8/3&0&28/3;-8&32/3;8&0&16/3;4&44/3;-4\cr
\hline\hline
\end{tabular}
}
\caption{\protect\label{tab2}Spin coefficients of the diagrams of fig.
\protect\ref{fig3}}
\end{table}

\subsection{Diagram 3}

The spin and isospin parts of the traces have the same structure as for 
the diagrams 1 and 2. The isospin coefficients are listed in table 
\ref{tab3} while the spin coefficients are listed in table \ref{tab4}.

\begin{table}
\centerline{
\begin{tabular}{||c||c|c|c|c||}
\hline\hline
~&SS&SV&VS&VV\cr
\hline\hline
a&2&6&2&-2\cr
\hline
b,c,d,e&0&0&0&8/3\cr
\hline
f,j&0&0&4/3&20/3\cr
\hline
g,i&0&0&0&4/9\cr
\hline
h,k&0&4&0&20/3\cr
\hline
l,m,n,o&0&0&0&-40/3\cr
\hline\hline
\end{tabular}
}
\caption{\protect\label{tab3} Isospin coefficients of the diagrams of 
fig. \protect\ref{fig4}}
\end{table}

\begin{table}
\centerline{
\begin{tabular}{||c||c|c|c|c|c|c|c|c|c||}
\hline\hline
~&SS&SL&ST&LS&LL&LT&TS&TL&TT\cr
\hline\hline
a&2&2&4&2&-2;4&0;-4&2&0;-2&-2;2\cr
\hline
b,c,d,e&0&0&0&0&2/3;2/3&2;-2/3&0&1;-1/3&5/3;1/3\cr
\hline
f,j&0&0&0&1/3&5/3;-1/3&4;4/3&4/3&2;2/3&14/3;-2/3\cr
\hline
g,i&0&0&0&0&-2/9;10/9&2/3;-10/9&0&1/3;-5/9&1/9;5/9\cr
\hline
h,k&0&4/3&8/3&0&5/3;-1/3&4;4/3&0&2;2/3&14/3;-2/3\cr
\hline
l,m,n,o&0&0&0&0&-22/3;26/3&-6;-26/3&0&-3;-13/3&-31/3;13/3\cr
\hline
g&0&4/3&8/3&0&28/3;-8&32/3;8&0&16/3;4&44/3;-4\cr
\hline\hline
\end{tabular}
}
\caption{\protect\label{tab4}Spin coefficients of the diagrams of 
fig. \protect\ref{fig4}}
\end{table}

\subsection{Diagrams 4 and 5}

In these diagrams spin-isospin traces have 3 components. As fig. 
\ref{fig5} clearly explain, the diagrams are obtained by connecting two 
fermionic loops (in this subsection we shall call them subdiagrams): we 
have to take the spin-isospin traces of the two subdiagrams, that depend 
upon the spin-isospin character 
of the probe and of the exchanged particles (first two 
factors), the traces however may generate two tensor in the isospin or 
configuration space, to be then saturated together, thus giving rise to 
the third factor (we shall call it ``connecting factor'').

The spin-isospin traces of the subdiagrams depend only upon the scalar 
or vector character of the probe and of the exchanged particles as far as the 
numerical coefficient is concerned, and amount to evaluate the traces of 
two or three generalized Pauli matrices, listed in appendix \ref{appA}. 
The difference between 
spin-longitudinal and spin-transverse channel instead only affects the 
connecting factor. Thus in table \ref{tab3} we give the numerical 
coefficients in the various cases. 
\begin{table}
\centerline{
\begin{tabular}{||c||c|c|c|c|c|c|c||}
\hline\hline
~&a&b&c&d&e&f&g\cr
\hline\hline
SSS&2&0&0&0&0&0&0\cr
\hline
SVV&2&0&4/3&0&0&4/3&0\cr
\hline
VVS&2&4/3&0&0&0&0&4/3\cr
\hline
VSV&2&0&0&4/3&4/3&0&0\cr
\hline
VVV&2&-2/3&-2/3&/-2/3&10/3&10/3&10/3\cr
\hline\hline
\end{tabular}
}
\caption{\protect\label{tab5} Numerical coefficient of the traces of the 
subdiagrams of diagrams 4 and 5. In the first column the scalar (S) or 
vector (V) character of the probe, the first (upper) and second (lower) 
exchanged particle are reported in the given order. The channels not 
reported here have vanishing coefficients.}
\end{table}

The spin-isospin traces are thus obtained by multiplying the entries of 
table \ref{tab5} corresponding to the required subdiagrams by the 
connecting factors given in table \ref{tab7} for the isospin and in 
table \ref{tab7} for the spin. The tensors arising from the spatial part 
of the traces are also listed.
\begin{table}
\centerline{
\begin{tabular}{||c|c|c|c|c||}
\hline\hline
probe & upper part. & lower part. &tensor&coeff.\cr
\hline\hline
S&S&S&I&1\cr
\hline
S&S&V& &0\cr
\hline
S&V&S& &0\cr
\hline
S&V&V&$\delta_{ij}$&3\cr
\hline
V&S&S& &0\cr
\hline
V&S&V&$\delta_{i3}$&1\cr
\hline
V&V&S&$\delta_{i3}$&1\cr
\hline
V&V&V&$i\epsilon_{ij3}$&$\mp 2$\cr
\hline\hline
\end{tabular}
}
\caption{\protect\label{tab6} Connecting coefficients of the isospin 
traces.}
\end{table}

\begin{table}
\centerline{
\begin{tabular}{||c|c|c|c|c|c|c|c||}
\hline\hline
probe & upper part. & lower part. &tensor& ~ & ~ & ~ & ~  \cr
\hline\hline
S & S & S & 1 & 1 & 0 & 0 & 0 \cr
S & S & L & 0 & 0 & 0 & 0 & 0 \cr
S & S & T & 0 & 0 & 0 & 0 & 0 \cr
S & L & S & 0 & 0 & 0 & 0 & 0 \cr
S & L & L & $\delta_{ij}p_iq_j\delta_{kl}p_kq_l$ & 0 & 0 & 0 & 1 \cr
S & L & T & $\delta_{ij}p_i \delta_{kl}p_k \tilde q_{jl}$ & 1 & 0 & 0 & 
-1 \cr
S & T & S & 0 & 0 & 0 & 0 & 0 \cr
S & T & L & $\delta_{ij}q_j \delta_{kl}q_l \tilde p_{ik}$ & 1 & 0 & 0 & 
-1 \cr
S & T & T & $\delta_{ij} \delta_{kl} \tilde p_{ik} \tilde q_{jl}$ & 1 & 0 
& 0 & 1 \cr
L & S & S & 0 & 0 & 0 & 0 & 0 \cr
L & S & L & $\delta_{ij} k_i q_j \delta_{kl} k_k q_l$ & 0 & 1 & 0 & 0 \cr
L & S & T & $\delta_{ij} k_i \delta_{kl} k_k \tilde q_{jl}$ & 1 & -1 & 0 
& 0 \cr
L & L & S & $\delta_{ij} k_i p_j \delta_{kl} k_k p_l$ & 0 & 0 & 1 & 0 \cr
L & L & L & $i\epsilon_{ijk}k_ip_jq_k i \epsilon_{lmn}k_lp_mq_n$ & 0 & 0 & 
0 & 0 \cr
L & L & T & $i\epsilon_{ijk}k_ip_ji \epsilon_{lmn}k_lp_m \tilde q_{kn}$ & 
$\mp 1 $ & 0 & $\pm 1$ & 0 \cr
L & T & S & $\delta_{ij} k_i \delta_{kl} k_k \tilde p_{jl}$ & 1 & 0 & -1 
& 0 \cr
L & T & L & $i\epsilon_{ijk}k_i q_k i \epsilon_{lmn}k_l q_n \tilde p_{jm}$ 
& $\mp 1$ & $\pm 1$ & 0 & 0 \cr
L & T & T & $i\epsilon_{ijk}k_i i \epsilon_{lmn}k_l \tilde p_{jm} \tilde 
q_{ln}$ & 0 & $\mp 1$ & $\mp 1$ & 0 \cr
T & S & S & 0 & 0 & 0 & 0 & 0 \cr
T & S & L & $\delta_{ij} q_j \delta_{kl} q_l \frac{1}{2}\tilde k_{ik}$ & 
$\frac{1}{2}$ & $-\frac{1}{2}$ & 0 & 0 \cr
T & S & T & $\delta_{ij}  \delta_{kl}  \frac{1}{2}\tilde k_{ik} \tilde 
q_{jl}$ & $\frac{1}{2}$ & $\frac{1}{2}$ & 0 & 0 \cr
T & L & S & $\delta_{ij} p_j \delta_{kl} p_l \frac{1}{2}\tilde k_{ik}$
& $\frac{1}{2}$ & 0 & $-\frac{1}{2}$ & 0 \cr
T & L & L &  $i\epsilon_{ijk} p_j q_k i \epsilon_{lmn} p_l q_n 
\frac{1}{2}\tilde k_{ik}$ & $\mp\frac{1}{2}$ & 0 & 0 & $\pm\frac{1}{2}$ \cr
T & L & T & $ i\epsilon_{ijk} p_j i \epsilon_{lmn} p_m \frac{1}{2}\tilde 
k_{ik} \tilde q_{kn}$ & 0 & 0 & $\mp\frac{1}{2}$ & $\mp\frac{1}{2}$ \cr
T & T & S & $\delta_{ij} \delta_{kl} \frac{1}{2}\tilde k_{ik} \tilde 
p_{jl}$ & $\pm\frac{1}{2}$ & 0 & $\pm\frac{1}{2}$ & 0 \cr
T & T & L & $i\epsilon_{ijk} q_k i \epsilon_{lmn} q_n \frac{1}{2}\tilde 
k_{il} \tilde p_{jm}$ & 0 & $\mp\frac{1}{2}$ & 0 & $\mp\frac{1}{2}$ \cr
T & T & T & $i\epsilon_{ijk}i \epsilon_{lmn}\frac{1}{2}\tilde k_{il} 
\tilde p_{jm} \tilde q_{kn}$ & $\mp\frac{3}{2}$ & $\pm\frac{1}{2}$ &
$\pm\frac{1}{2}$ & $\pm\frac{1}{2}$ \cr
\hline\hline
\end{tabular}
}
\caption{\protect\label{tab7} Connecting factors of the spatial traces.
S, L and T refer respectively to scalar, spin-longitudinal and 
spin-transverse particles. The $\pm$ sign, when occurs, refers to 
diagram 4 (upper sign) and 5 (lower sign). The four last entries are 
respectively the constant term, the coefficient of $\cos^2\theta_{kq}$,
$\cos^2\theta_{kp}$ and $\cos^2\theta_{pq}$.}
\end{table}


\newpage

\begin{thebibliography}{10}

\bibitem{CeSa-94}
{R. Cenni and P. Saracco}.
\newblock {\em Phys. Rev.}, C50:1851, 1994.

\bibitem{Me-al-84}
{Z. E. Meziani et al.}
\newblock {\em Phys. Rev. Lett.}, 52:2180, 1984.

\bibitem{Me-al-85}
{Z. E. Meziani et al.}
\newblock {\em Phys. Rev. Lett.}, 54:1233, 1985.

\bibitem{BoGiPa-93}
{S. Boffi, C. Giusti and F. D. Pacati}.
\newblock {\em Phys. Rep.}, 226C:1, 1993.

\bibitem{Zg-al-94}
{A. Zghiche et al.}
\newblock {\em Nucl. Phys.}, A572:513, 1994.

\bibitem{Ya-al-93}
{T. C. Yates et. al.}
\newblock {\em Phys. Lett.}, 312B:382, 1993.

\bibitem{Jo-95}
{J. Jourdan}.
\newblock {\em Phys. Lett.}, B353:189, 1995.

\bibitem{Ba-al-94}
{M. B. Barbaro et al.}
\newblock {\em Nucl. Phys.}, B569:701, 1994.

\bibitem{Jo-96}
{J. Jourdan}.
\newblock {\em Nucl. Phys.}, A603:117, 1996.

\bibitem{FaPa-87}
{S. Fantoni and V. R. Pandharipande}.
\newblock {\em Nucl. Phys.}, A473:234, 1987.

\bibitem{CoQuSmWa-88}
{G. P. C\`o, K. F. Quader, R. D. Smith and J. Wambach}.
\newblock {\em Nucl. Phys.}, A485:61, 1988.

\bibitem{Ba-al-95}
{M. Barbaro et al.}
\newblock To be published on Nucl. Phys. A.

\bibitem{VaRyWa-95}
{V. Van der Sluys, J. Rickebusch and M. Waroquier}.
\newblock To be published on Phys. Rev. C.

\bibitem{AlErMo-84}
{W. M. Alberico, M. Ericson and A. Molinari}.
\newblock {\em Ann. Phys.}, 154:356, 1984.

\bibitem{AmCoLa-94}
{J. E. Amaro, G. C\`o and A. M. Lallena}.
\newblock {\em Nucl. Phys.}, A578:365, 1994.

\bibitem{CeFa-93}
{R. Cenni and S. Fantoni}.
\newblock {\em Phys. Lett.}, B314:260, 1993.

\bibitem{WaCeFaFa-96}
{T. S. Walhout, R. Cenni, A. Fabrocini and S. Fantoni}.
\newblock {\em Phys. Rev.}, C54:1622, 1996.

\bibitem{Fa-96-t}
{A. Fabrocini}.
\newblock "To be published on the proceedings of the ``Workshop on electron
  nucleus scattering'' - Marciana Marina, Elba, July 1-5 1996".

\bibitem{CeCoSa-96}
{R. Cenni, F. Conte and P. Saracco}.
\newblock {\em J. Phys. G}, 22:L71, 1996.

\bibitem{AlCeMoSa-87}
{W. M. Alberico, R. Cenni, A. Molinari and P. Saracco}.
\newblock {\em Ann. of Phys.}, 174:131, 1987.

\bibitem{Ce-90}
{R. Cenni}.
\newblock volume~6 of {\em Condensed Matter Theories}.
\newblock Plenum Press, New York and London, 1990.
\newblock Edited by S. Fantoni and S. Rosati.

\bibitem{AlCeMoSa-88}
{W. M. Alberico, R. Cenni, A. Molinari and P. Saracco}.
\newblock {\em Phys. Rev.}, C38:2389, 1988.

\bibitem{CeSa-88}
{R. Cenni and P. Saracco}.
\newblock {\em Nucl. Phys.}, A487:279, 1988.

\bibitem{CeCoCoSa-92}
{R. Cenni, F. Conte, A. Cornacchia and P. Saracco}.
\newblock {\em La Rivista del Nuovo Cimento}, {15 n. 12}, 1992.

\bibitem{AlCeMoSa-90}
{W. M. Alberico, R. Cenni, A. Molinari and P. Saracco}.
\newblock {\em Phys. Rev. Lett.}, 65:1845, 1990.

\bibitem{MoGaNaRa-74}
{G. Morandi, E. Galleani d'Agliano, F. Napoli and C. Ratto}.
\newblock {\em Adv. Phys.}, 23:867, 1974.

\bibitem{Ke-70}
{H. Keiter}.
\newblock {\em Phys. Rev.}, 2B:3777, 1970.

\bibitem{Kl-78}
{H. Kleinert}.
\newblock {\em Fort. Phys.}, 26:565, 1978.

\bibitem{SpWeWi-77}
{J. Speth, E. Werner and J. Wild}.
\newblock {\em Phys. Rep.}, 33:127, 1977.

\bibitem{Ce-93}
{R. Cenni}.
\newblock In {J. Arvieux and E. De Sanctis}, editor, {\em The ELFE project: an
  Electron Laboratory for Europe p. 437 -- Mainz, October 1992}, page 437,
  1993.

\bibitem{MaHoEl-87}
{R. Machleidt, K. Holinde and Ch. Elster}.
\newblock {\em Phys. Rep.}, C149:1, 1987.

\bibitem{BrBaOsWe-77}
E.~Oset G.~E.~Brown, S. O.~B\"ackman and W.~Weise.
\newblock {\em Nucl. Phys.}, A286:191, 1977.

\bibitem{OsSa-87}
{E. Oset and L. L. Salcedo}.
\newblock {\em Nucl. Phys.}, A468:631, 1987.

\bibitem{OsWe-79}
{E. Oset and W. Weise}.
\newblock {\em Nucl. Phys.}, A319:365, 1979.

\bibitem{OsWe-79a}
{E. Oset and W. Weise}.
\newblock {\em Nucl. Phys.}, A329:47, 1979.

\bibitem{BrWe-75}
{G. E. Brown and W. Weise}.
\newblock {\em Phys. Rep.}, C22:281, 1975.

\bibitem{De-84}
{T. De Forest, jr.}
\newblock {\em Nucl. Phys.}, A414:347, 1984.

\bibitem{Gi-90}
{M. M. Giannini}.
\newblock {\em Rep. Prog. Phys.}, 54:453, 1990.

\bibitem{ErRo-86}
{M. Ericson and M. Rosa-Clot}.
\newblock {\em Zeit. Phys.}, A324:373, 1986.

\bibitem{AlCeMo-89}
{W. M. Alberico, R. Cenni, and A. Molinari}.
\newblock {\em Prog. in Part. and Nucl. Phys.}, 23:171, 1989.

\bibitem{Wa-74}
{J. D. Walecka}.
\newblock {\em Ann. Phys.}, 83:491, 1974.

\bibitem{ClHaMe-82}
{B. C. Clark, S. Hama and R. L. Mercer}.
\newblock In H.~O. Meyer, editor, {\em The Interaction Between Medium Energy
  Nucleons in Nuclei}, page 260. AIP Conference Proceedings, 1982.

\bibitem{Cl-86}
{B. C. Clark}.
\newblock In {M. B. Johnson and A. Picklesimer}, editor, {\em Relativistic
  Dynamics and Quark-Nuclear Physics}, page 302, New York, 1986. J. Wiley \&
  Sons.

\bibitem{Ho-81}
{K. Holinde}.
\newblock {\em Phys. Rep.}, C68:121, 1981.

\bibitem{DaMcDoSi-90}
{D. B. Day, J. S. McCarthy, T. W. Donnelly and I. Sick}.
\newblock {\em Ann. Rev. Nucl. Part. Sci.}, 40:357, 1990.

\bibitem{CaOs-92}
{R. C. Carrasco and E. Oset}.
\newblock {\em Nucl. Phys.}, A536:445, 1992.

\bibitem{CeDoMo-97-t}
{R. Cenni, T. W. Donnelly and A. Molinari}.
\newblock Submitted to Phys. Rev. C.

\end{thebibliography}
\end{document}